\RequirePackage[immediate]{silence}
\documentclass[journal]{IEEEtran}
\usepackage{amsfonts}
\usepackage{amssymb} 
\usepackage{amsmath}
\usepackage{dsfont}
\usepackage{lipsum}
\usepackage{graphicx}
\usepackage{cite}    
\usepackage{bigstrut}
\usepackage{xcolor,colortbl}
\usepackage{float}
\usepackage{balance}
\usepackage{verbatim}
\usepackage{lipsum}
\usepackage{psfrag}
\usepackage{multirow}
\usepackage{bbm}
\usepackage{multicol}

\usepackage{enumitem}
\newtheorem{theorem}{Theorem}

\newtheorem{corollary}{Corollary}

\newtheorem{assumption}{Assumption}

\newtheorem{remark}{Remark}

\usepackage{alphabeta}
\usepackage{amsmath,amssymb}
\usepackage{verbatim}
\usepackage{bookmark}
\usepackage{float}
\usepackage{algpseudocode}
\usepackage{algorithm}
\usepackage{mathtools, nccmath}
\usepackage{subcaption}
\newcommand\norm[1]{\left\lVert#1\right\rVert}

\newcommand{\inprod}[1]{\left\langle #1 \right\rangle}
\definecolor{lightgray}{rgb}{0.75,0.75 ,0.75 }
\newcolumntype{P}[1]{>{\centering\arraybackslash}p{#1}}
\newcommand*{\Scale}[2][4]{\scalebox{#1}{$#2$}}

\begin{document}
	\IEEEoverridecommandlockouts
	
	\title{Accelerating Distributed Optimization via Over-the-Air Computing}
	\author{Nikos A. Mitsiou, \IEEEmembership{Student Member, IEEE}, Pavlos S. Bouzinis, \IEEEmembership{Student Member, IEEE},\\ Panagiotis D. Diamantoulakis,~\IEEEmembership{Senior Member, IEEE}, Robert Schober, \IEEEmembership{Fellow, IEEE}, \\ George K. Karagiannidis,~\IEEEmembership{Fellow, IEEE   \vspace{-1cm}}
\thanks{}
\thanks{N.~A.~Mitsiou, P.~S.~Bouzinis, P.~D.~Diamantoulakis, and G.~K.~Karagiannidis are with the Wireless Communications and Information Processing (WCIP) Group,  Department of Electrical and Computer Engineering, Aristotle University of Thessaloniki, 54636, Thessaloniki, Greece \, (e-mail: nmitsiou@ece.auth.gr, mpouzinis@ece.auth.gr, padiaman@ieee.org, geokarag@auth.gr).}
\thanks{R. Schober is with Friedrich-Alexander-University Erlangen-Nürnberg
 (FAU), 91058 Erlangen, Germany, e-mail: robert.schober@fau.de}
\vspace{-0cm}}
	\maketitle
\vspace{-0.3in}
\begin{abstract} 
Distributed optimization is ubiquitous in emerging  applications, such as robust sensor network control, smart grid management, machine learning, resource slicing, and localization. However, the extensive data exchange among local and central nodes may cause a severe communication bottleneck. To overcome this challenge, over-the-air computing (AirComp) is a promising medium access technology, which exploits the superposition property of the wireless multiple access channel (MAC) and offers significant bandwidth savings. In this work, we propose an AirComp framework for general distributed convex optimization problems. Specifically, a distributed primal-dual (DPD) subgradient method is utilized for the optimization procedure.  Under general assumptions, we prove that DPD-AirComp can asymptotically achieve  zero expected constraint violation. Therefore, DPD-AirComp ensures the feasibility of the original problem, despite the presence of channel fading and additive noise. Moreover, with proper power control of the users’ signals, the expected non-zero optimality gap can also be mitigated. Two practical applications of the proposed framework are presented, namely, smart grid management and wireless resource allocation. Finally, numerical results reconfirm DPD-AirComp's excellent performance, while it is also shown that DPD-AirComp converges an order of magnitude faster compared to a digital orthogonal multiple access scheme, specifically, time division multiple access (TDMA). 
\end{abstract}

\section{Introduction}
Distributed optimization  has drawn considerable attention for addressing a plethora of problems in various fields, ranging from distributed machine learning to resource allocation and optimization of wireless networks \cite{comp1,comp2}. Distributed optimization refers to the idea of leveraging the computational power of multiple devices/agents to solve an optimization problem efficiently. Thus, to enable the implementation of distributed optimization, a  problem  needs to be decomposed into disjoint subproblems of smaller size, and each agent is assigned  a specific subproblem. In the context of wireless networks, distributed optimization can be realized by allowing all the network's participants, i.e., the central coordinator and the  devices in the underlying physical layer, to actively collaborate towards obtaining a global solution, facilitating the practical implementation of solutions that are based on optimization in large networks \cite{6gbook}.

Considering its benefits, distributed optimization is expected to be one of the enablers of future wireless networks in the sixth generation (6G)  era \cite{6g1}, \cite{6g2}.
This is due to the fact that 6G will aim to incorporate intelligence into the physical (PHY) and the medium-access control layers the network, eventually, giving rise to new functionalities, such as intelligent power control, intelligent interference management, and joint optimization \cite{6gbook}. Traditional centralized algorithms are an option for making those intelligent decisions, however, they can be easily overwhelmed by the increasing number of users and the huge amounts of data. Due to its inherent scalability, distributed optimization can overcome this challenge. Nonetheless, distributed optimization requires wireless transmission of big volumes of data between the participants, which  might cause communication bottlenecks and significant transmissions overheads in large-scale networks.

In this direction, non-orthogonal protocols, such as power-domain non-orthogonal multiple access (NOMA) and rate-splitting multiple access (RSMA), where interference is decoded, rather than suppressed, can be utilized to increase spectral efficiency  \cite{bouz1, bouz2}. Although breaking orthogonality in the downlink has been well-investigated in the existing literature, transmitting information in the uplink by using the same resources in the time, frequency, and code domains has not been thoroughly investigated, despite its high practical value for a vast number of wireless applications, including the Internet of Things. Moreover, when the aim is to compute a certain function of the distributed data, a promising non-orthogonal technology, which aims to alleviate the burden of increased congestion and communication traffic, is over-the-air computing (AirComp) \cite{ota1,ota2,ota3,ota4,ota5}. The core idea of AirComp is to exploit the inherent wave superposition property of the multiple access channel (MAC) towards computing a desired function. Specifically, the devices’ messages are simultaneously transmitted over the MAC and aggregated “over-the-air” when arriving at the fusion center. Notably, by appropriate preprocessing, AirComp enables the computation of nomographic functions, and thus any other function \cite{ota3}. As a result, AirComp  has the potential to accomplish ultra-fast data aggregation, and thus provides a simple but effective protocol, upon which distributed optimization for future wireless networks can be efficiently implemented.
\vspace{-0.2cm}
\subsection{Related Works}
\vspace{-0.1cm}
Distributed  optimization has been extensively studied over the years \cite{opt1,opt2,opt3,opt4,opt5,opt6,opt7}. In \cite{opt1}, distributed optimization was investigated for solving nonconvex optimal power flow problems in real large-scale power systems. Moreover, in \cite{opt2}, an event-driven distributed optimization scheme was developed in order to reduce communication costs, with application to sensor network coverage control. The impact of communication delay on distributed optimization was  examined in \cite{opt3}, where Lyapunov theory for time delay systems was used and two illustrative examples were presented. Furthermore, in \cite{opt4}, the effect of limited capacity communication links on various distributed subgradient algorithms was studied, while a dimensionality reduction mechanism was proposed. In addition, in \cite{opt5}, the convergence of the dual subgradient averaging method was analyzed, in the context of distributed optimization, and the impact of wireless communication was studied. Finally, in \cite{opt6,opt7}, achieving consensus through distributed optimization was examined. Specifically, in \cite{opt6}, the alternating direction method of multipliers was adopted, while in \cite{opt7}, a lazy mirror descent method was designed and the impact of limited channel capacity and noise were investigated.

To render distributed optimization more communication efficient, various works have leveraged the concept of AirComp to enable low-latency MAC communication. However, AirComp has been extensively studied only in the context of federated learning (FL) \cite{fl2,fl3,fl4,fl5,fl7,fl8,fl9,fl10,fl11,fl12}. For instance, in \cite{fl2}, an AirComp FL scheme was developed aiming to cope with the heterogeinity of the participating users.
Also, in \cite{fl5}, an analog FL framework was proposed in which the devices first sparsify their gradient estimates and project the resulting sparse vectors into low-dimensional vectors before transmitting them to a central unit. Moreover, in \cite{fl7,fl11}, a joint device selection and beamforming design was studied for improving over-the-air FL, while in \cite{fl9} NOMA and intelligent reflective surfaces were integrated into AirComp FL. In addition, in \cite{fl3,fl4}, a device-to-device (D2D) AirComp FL scheme was proposed, and the impact of the communication between all nodes was studied. Furthermore, in \cite{fl8,fl10} the joint optimization of power control and FL in an AirComp scenario was considered, while in \cite{fl12}, a  parallel FL framework based on AirComp was proposed with joint receiver-combiner vector design and device selection.  Finally, in \cite{dota}, an AirComp dual-averaging framework was formulated, aiming to solve an optimization problem in which all devices share a common objective function, which is the average of all local devices' objective functions.    
\vspace{-0.2in}
\subsection{Motivation and Contributions}
Most existing works \cite{fl2,fl3,fl4,fl5,fl7,fl8,fl9,fl10,fl11,fl12}, which applied the AirComp concept to distributed optimization problems, focused on federated and distributed learning techniques. However, the latter approaches may not adequately capture the structure of a wide range of optimization problems. For instance, in FL, local solvers are not  subject to additional local or global constraints, while the global objective function is usually separable, and given by the sum of the local functions. Also, differentiable local functions are usually assumed. On the other hand, the application of AirComp to distributed optimization problems with a more general structure, e.g., constrained non-differentiable problems has yet to be examined.
\par
An efficient approach to solve a general, non-differentiable, constrained optimization problem with non-separable objective function in a distributed manner is the primal-dual subgradient method. Primal-dual subgradient optimization \cite{nedic1,nedic2} can be applied to a variety of problems and in many different fields, including wireless communications. It is beneficial for solving large-scale optimization problems where a large number of variables may render the problem intractable, e.g., when second-order methods are adopted. Moreover, Lagrangian duality is an effective tool for  providing lower bounds on the optimal value of nonconvex optimization problems, which often arise in wireless communication applications. 

\par 
Driven by the aforementioned considerations, we investigate the use of AirComp as uplink multiple access protocol, to solve optimization problems with general structure in a distributed manner. Specifically, we adopt the distributed primal-dual subgradient method, where users utilize  AirComp  to convey the dual variables to a central server, whose responsibility it is to update the primal variables and broadcast them back to the users. After designing and presenting the proposed distributed primal dual AirComp (DPD-AirComp) algorithm, we rigorously examine its convergence behavior. Finally, we evaluate the performance of the proposed DPD-AirComp algorithm both in terms of optimality and convergence speed, 
by applying  it to two practical optimization scenarios, namely, smart grid management and wireless resource allocation. The contributions of this paper can be summarized as follows:
\begin{itemize}[leftmargin=*]
    \item A novel AirComp framework for distributed primal-dual optimization is proposed, namely DPD-AirComp, which can be used for solving a wide-range of distributed optimization problems, while promoting communication efficiency and guaranteeing high convergence speed. Specifically, we adopt a distributed primal-dual subgradient method, where users utilize  AirComp  to convey the dual variables to the central server, which subsequently updates the primal variables and broadcasts them back to the users.
    \item The convergence  of the DPD-AirComp algorithm is analyzed. A key point of the analysis is the partial user participation throughout the iterative process of the algorithm, which is a result of  AirComp's principles. Notably, it is shown that the expected constraint violation of the optimization problem tends to zero asymptotically, highlighting that AirComp does not affect the feasibility of the underlying problem. However, the partial user participation creates a non-zero optimality gap, which, though, can be mitigated by properly preprocessing the users' transmit signals.
    \item The proposed DPD-AirComp algorithm is applied to two distributed optimization use cases. Specifically, a smart grid energy management system and a frequency-division multiple access (FDMA) distributed resource allocation problem are considered. The proposed algorithm is analyzed and tested, given the particularities of each considered scenario. Simulations are conducted to evaluate the performance of the proposed DPD-AirComp algorithm. The results verify the effectiveness of the proposed method, highlighting its near-optimal performance and the significant convergence time acceleration of DPD-AirComp compared to the conventional time-division multiple access (TDMA) protocol. 
\end{itemize}
\vspace{0cm}
\subsection{Notation} 
$\mathbb{R}$ represents the set of real numbers, while $\mathbb{R}_{+}$ represents the set of real positive numbers.
$|\cdot|$ denotes the cardinality of a set or the absolute value of a number, depending on the respective context. Bold characters denote vectors.   $\inprod{\boldsymbol{x},\boldsymbol{y}}$ denotes the inner product of two vectors of equal dimension, $\boldsymbol{x}$ and $\boldsymbol{y}$. Moreover, $\norm{\cdot}$ represents the standard Euclidean norm and $\mathbb{E}[\cdot]$ denotes  expectation. $\succeq$ denotes an element-wise vector inequality, while $[\cdot]^+$ stands for $\mathrm{max}\{0,\cdot\}$.                                                                                

\section{Problem Statement and AirComp Implementation}
In this section, the proposed system architecture is introduced. First, in subsection II.A, we provide background information on DPD optimization. Then, in subsection II.B, the AirComp implementation of DPD optimization is discussed, which results in a distributed optimization AirComp framework, namely, DPD-AirComp.
\subsection{Distributed Primal-Dual Optimization}
We consider a wireless network consisting of $N$ devices, indexed by $i\!\in \! \mathcal{N}\!=\!\{1,2,...,N\}\!$ and a base station (BS) collocated with a central server. Both the devices and the BS are equipped with a single antenna. We assume that the central server is interested in minimizing a global objective function $f_0:\mathbb{R}^D \rightarrow \mathbb{R}$, subject to the local constraints associated with each individual device, and global constraints which have to be met by all devices and/or the BS.
The considered problem can be formally written as follows
\begin{equation} \label{eq:original}
\begin{aligned}
\underset{{\boldsymbol{x}}}{\text{min}} \quad &f_0\left(\boldsymbol{x}\right) \\
\quad \text{s.t.}\quad   &f_i\left(\boldsymbol{x}\right)\leq 0, \,\,\, i \in \mathcal{N}, \\
\quad &\boldsymbol{x}\in \mathcal{X},
\end{aligned}
\end{equation}
where the functions $f_0,f_1,..,f_N:\mathbb{R}^D \rightarrow \mathbb{R}$ are convex, not necessarily differentiable, and indicate the devices' local constraints. Also, $\mathcal{X} \subset \mathbb{R}^D$ is a nonempty, compact, and convex set, reflecting the set of global constraints.
\par
We aim to solve the problem in \eqref{eq:original} in a distributed manner.
A common approach for solving such a problem is the distributed primal-dual method \cite{nedic1,nedic2}. Specifically, the dual problem of \eqref{eq:original} is defined through the Lagrangian relaxation of the inequalities constraints, and can be written as follows
\begin{equation} 
\begin{aligned}
\underset{\boldsymbol{\lambda}}{\text{max}} \quad & q\left(\boldsymbol{\lambda}\right)\\
\quad \text{s.t.}\quad &\boldsymbol{\lambda} \succeq 0, \,\\
&\boldsymbol{\lambda} \in \mathbb{R}^{N},
\end{aligned}
\end{equation}
where the dual objective function is defined as 
\begin{equation} \label{dual_objective}
q\left(\boldsymbol{\lambda}\right)= \underset{\boldsymbol{x} \in \mathcal{X}}{\text{inf}} \mathcal{L}\left(\boldsymbol{x},\boldsymbol{\lambda}\right).
\end{equation}
In \eqref{dual_objective}, $\mathcal{L}\left(\boldsymbol{x},\boldsymbol{\lambda}\right): \mathcal{X}\times \mathbb{R}_{+}^N \rightarrow \mathbb{R} $ is the Lagrangian function defined as 
\begin{equation}  
\mathcal{L}\left(\boldsymbol{x},\boldsymbol{\lambda}\right) = f_0\left(\boldsymbol{x}\right)+\sum_{i=1}^N \lambda_if_i\left(\boldsymbol{x}\right),
\end{equation}
where $\lambda_i$ is the Lagrange multiplier associated with the $i$-th constraint. Since we do not assume differentiability of  functions $f_0,f_i, \, \forall i \in \mathcal{N}$, we define
\begin{equation}  
\begin{aligned}\label{subgradients}
&\boldsymbol{\mathcal{L}}_x\left(\boldsymbol{x},\boldsymbol{\lambda}\right)=\boldsymbol{g}_0\left(\boldsymbol{x}\right)+\sum_{i=1}^N \lambda_i \boldsymbol{g}_i\left(\boldsymbol{x}\right),\,\, \text{and}\,\,\, \\
&\boldsymbol{\mathcal{L}}_\lambda\left(\boldsymbol{x},\boldsymbol{\lambda}\right)=\boldsymbol{F}\left(\boldsymbol{x}\right),
\end{aligned} 
\end{equation}
to denote the subgradients of $\mathcal{L}\left(\boldsymbol{x},\boldsymbol{\lambda}\right)$ with respect to (w.r.t.) $\boldsymbol{x}$ and $\boldsymbol{\lambda}$, where $\boldsymbol{g}_0,\boldsymbol{g}_1,...,\boldsymbol{g}_N$ denote the subgradients of functions $f_0,f_1,...,f_N$, respectively, and $\boldsymbol{F}\left(\boldsymbol{x}\right)=\left[f_1\left(\boldsymbol{x}\right),...,f_N\left(\boldsymbol{x}\right)\right]^T$. Specifically, $\boldsymbol{g}(\boldsymbol{x}')\in \mathbb{R}^D$ is a subgradient of a convex function $f:\mathbb{R}^D \rightarrow \mathbb{R}$ for a given vector $\boldsymbol{x}' \in \mathcal{X}$, when the following relation holds:
\begin{equation}  
    f(\boldsymbol{x}')+\inprod{\boldsymbol{g}(\boldsymbol{x}'),\boldsymbol{x}-\boldsymbol{x}'}\leq f(\boldsymbol{x}).
\end{equation}
The primal-dual algorithm iteratively updates the primal and dual variables with the aid of the sub-gradient method. Specifically, in the $k$-th iteration/round, the variables are updated as follows:  
\begin{equation}  \label{x}
\begin{aligned}
\boldsymbol{x}^{k+1}&= \mathcal{P}_\mathcal{X}\left[ \boldsymbol{x}^{k}-a_k\boldsymbol{\mathcal{L}}_x\left(\boldsymbol{x}^k,\boldsymbol{\lambda}^k\right) \right]\\
&=\mathcal{P}_\mathcal{X}\left[ \boldsymbol{x}^{k}-a_k\left( \boldsymbol{g}_0\left(\boldsymbol{x}^k\right)+\sum_{i=1}^N\lambda^k_i\boldsymbol{g}_i\left(\boldsymbol{x}^{k}\right)\right) \right]
\end{aligned}
\end{equation}
and
\begin{equation}  
\begin{aligned}
\boldsymbol{\lambda}^{k+1}&=\mathcal{P}_\mathcal{D}\left[\boldsymbol{\lambda}^k+a_k\boldsymbol{\mathcal{L}}_{\lambda}\left(\boldsymbol{x}^k,\boldsymbol{\lambda}^k\right)\right]\\
&=\mathcal{P}_\mathcal{D}\left[\boldsymbol{\lambda}^k+a_k\boldsymbol{F}\left(\boldsymbol{x}^{k}\right)\right],
\end{aligned}
\end{equation}
where $a_k \in (0,1)$ is the stepsize of the $k$-th round, while $\mathcal{P}_{\mathcal{S}}\left[\cdot\right]$ denotes the projection operator onto set $\mathcal{S}$ and it is defined as
\begin{equation}  
    \mathcal{P}_{\mathcal{S}}\left[z\right]=\underset{s\in\mathcal{S}}{\arg\!\min}\norm{s-z}^2.
\end{equation}
Set $\mathcal{D}$ contains the Lagrange multipliers, and it is compact and convex. According to \cite{nedic1}, under the Slater condition, dual variable $\boldsymbol{\lambda}$ is bounded, and thus, an appropriate choice of $\mathcal{D}$ is given as follows \cite{nedic1}
\begin{equation}  
\mathcal{D} = \bigg\{\boldsymbol{\lambda} \succeq \boldsymbol{0} \, \bigg\vert \, \norm{\boldsymbol{\lambda}}_{\infty} \leq \frac{f_0\left(\boldsymbol{\bar{x}}\right)-\tilde{q}}{\gamma} + r \bigg\},  
\end{equation}
where $\boldsymbol{\bar{x}}$ is a vector satisfying the Slater condition, $\tilde{q}=q(\boldsymbol{\lambda})$, for any $\boldsymbol{\lambda}\succeq 0$, $\gamma=\underset{1\leq i\leq N}{\mathrm{min}}\{-f_i(\boldsymbol{\bar{x}})\}$ and $r$ is any value from $\mathbb{R}_+$. We note that the projection onto set $\mathcal{D}$ can be executed individually by each user, since the operation $\norm{\boldsymbol{\lambda}}_{\infty}\leq  \frac{f_0\left(\boldsymbol{\bar{x}}\right)-\tilde{q}}{\gamma} + r$ can be equivalently  written as $\lambda_i \leq \frac{f_0\left(\boldsymbol{\bar{x}}\right)-\tilde{q}}{\gamma} + r,$ $\forall i \in \mathcal{N}$. Therefore, the $i$-th user does not need feedback from the other users regarding the value of their Lagrange multipliers $\lambda_{{j\neq i},j}$, which enables the distributed calculation of the dual variables. Therefore, each individual device calculates
\begin{equation}  
\lambda_i^{k+1}=\mathcal{P}_\mathcal{D}\left[\lambda_i^k+a_kf_i\left(\boldsymbol{x}^{k}\right)\right], \quad \forall i \in\mathcal{N}.
\end{equation}
In contrast to this, the primal variable $\boldsymbol{x}$ is updated at the server in a centralized fashion, according to (7), after having collected the dual-variable updates from all devices in iteration $k$. 


\subsection{Over-the-Air Implementation}

As mentioned in the previous subsection, the central server and the devices aim to collaboratively solve the optimization problem in (1), with the aid of the distributed primal-dual method. The collaboration between the BS and the users is organized in $K$ communication rounds, where each communication round includes the uplink data aggregation at the BS and the downlink broadcast from the BS to the users. To that end, the BS first broadcasts the vector $\boldsymbol{x}^k$ to all devices during the $k$-th round. Following that, each individual device calculates $\lambda_i^{k+1}$ and afterwards transmits $\lambda_i^k\boldsymbol{g}_i(\mathbf{x}^k)$ to the BS. Finally, the server updates the primal variable $\boldsymbol{x}^{k+1}$ and the process is repeated until convergence. As one can observe from \eqref{x}, the server is not actually interested in acquiring each devices' $\lambda_i^k$ separately, but instead, it requires only knowledge of the sum $\sum_{i \in \mathcal{N}}\lambda^k_i\boldsymbol{g}_i\left(\boldsymbol{x}^{k}\right)$, across all devices. Driven by this observation,
we propose to adopt the concept of AirComp for the uplink message transmission of the devices to the BS. The core idea of AirComp is to exploit the wave superposition property of the (MAC).  As a result, the devices' messages, i.e., $\lambda^k_i\boldsymbol{g}_i\left(\boldsymbol{x}^{k}\right), \, \forall i \in \mathcal{N}$,  are simultaneously transmitted and aggregated ``over-the-air" at the BS. Hence, the received aggregated signal at the BS, in round $k$, can be written as
\vspace{-.0cm}
\begin{equation}  
\boldsymbol{y}^k=\sum_{i=1}^N h_i^k\boldsymbol{w}_i^k+\boldsymbol{n}^k,
\end{equation}
where $\boldsymbol{w}_i^k\in \mathbb{R}^D$ is the transmitted signal and $h^k_i$ the channel coefficient of the $i$-th user in the $k$-th round. Here, the uplink communication is divided into $D$ symbol slots, corresponding to the size of  vector $\boldsymbol{w}_i^k$. Moreover 
the channel coefficients are assumed to be quasi-static, i.e., static in a single round, but varying from one round to the next. Also, perfect local channel state information (CSI) is assumed to be available at all $N$ transmitters, which can be acquired with the aid of the BS via pilot symbols.
 
Assuming a peak power constraint, $P_{\mathrm{max}}$, for each user, we have
\begin{equation}  
\norm{\boldsymbol{w}_i^k}^2 \leq P_{\mathrm{max}}, \quad \forall i \in \mathcal{N}.
\end{equation}
In order to mitigate the destructive effects of fading, a channel inversion strategy is adopted \cite{fl2}. As such, each user sets 
\begin{equation}  
\boldsymbol{w}_i^k= \left\{
\begin{array}{ll} \label{dikladi}
			\frac{\boldsymbol{s}_i^{k}}{\sqrt{\beta} h_i^k}, \quad \norm{\frac{\boldsymbol{s}_i^{k}}{h_i^k}}^2 \leq \beta P_{\mathrm{max}} \\
			\quad 0  , \quad\quad \mathrm{otherwise}\\
\end{array}, 
\right.
\end{equation}
where $\boldsymbol{s}_i^{k}=\lambda_i^k\boldsymbol{g}_i^k(\boldsymbol{x}^k)$ and $\beta$ is a preprocessing scalar to guarantee that all users meet their power constraint in (13). Specifically, its purpose lies in regulating the power of the transmitted signal (14) of all users, so that (13) is satisfied.  

Furthermore, due to the power constraint of the users' transmitters and the channel inversion strategy, only a subset of users will participate in any given communication round. We make the assumption that if there exists a bottleneck device with severe path-loss for all transmission rounds, which does not allow channel inversion within the given power budget, this device will be excluded from the optimization procedure \cite{one-bit}. Thus, the local constraints of that device are not taken into account. For all other devices, by using  (14), and the power constraint in (13), we  have 
\begin{equation}  
\norm{\boldsymbol{w}_i^k}^2 \leq P_\mathrm{max} \Leftrightarrow|h_i^k|^2 \geq \frac{\norm{\boldsymbol{s}_i^k}^2}{\beta P_\mathrm{max}},
\end{equation}
which reflects the condition for a user to participate. By taking this into account, the subset of  participating users in the $k$-th round is given by
\begin{equation}  
\mathcal{A}^k\triangleq \left\{i \in \mathcal{N} \,\, \bigg\vert \,\, |h_i^k|^2 \geq \frac{\norm{\boldsymbol{s}_i^k}^2}{\beta P_\mathrm{max}} \right\} \subseteq \mathcal{N}, 
\end{equation}
with $\lvert\mathcal{A}^k\rvert=A^k$. Notice here that the number of  participating users may differ from round to round, due to small-scale fading variations. The probability that the $i$-th user participates during the $k$-th round is given by
\begin{equation}  
\gamma_i^k\left(\beta\right)=\mathrm{Pr}\left\{|h_i^k|^2 \geq \frac{\norm{\boldsymbol{s}_i^k}^2}{\beta P_\mathrm{max}}\right\}, \quad \forall i \in \mathcal{N}.
\end{equation}
Finally, according to (12), the received signal at the BS is be given as
\begin{equation}  
\boldsymbol{y}^k=\sum_{i \in \mathcal{A}^k}\frac{\boldsymbol{s}_i^{k}}{\sqrt{\beta}}+\boldsymbol{n}^k.
\end{equation} 

\begin{figure}[t!]
\vspace{-0cm}
\centering
\includegraphics[width=\linewidth]{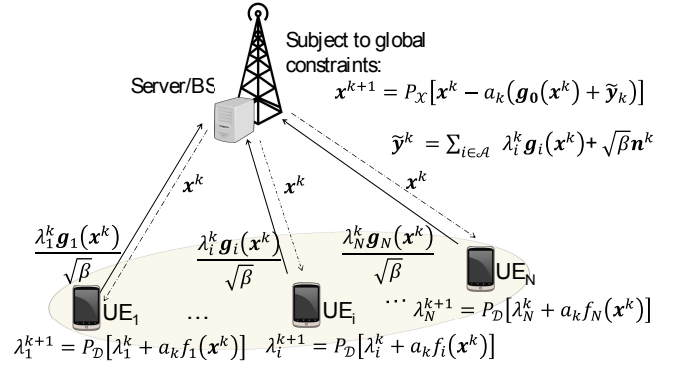}
\centering
\vspace{-0.5cm}
\caption{System model.}  \label{fig:fig0}
\vspace{-1cm}
\end{figure}

In order to recover the desired signal, $\sum_{i \in \mathcal{N}} \boldsymbol{s}_i^{k}$, the BS multiplies the received signal with $\sqrt{\beta}$ and obtains
\begin{equation}  
\tilde{\boldsymbol{y}}^k=\sum_{i \in \mathcal{A}^k}\boldsymbol{s}_i^{k}+\sqrt{\beta}\boldsymbol{n}^k.
\end{equation}
We note that the received signal, $\tilde{\boldsymbol{y}}^k$, is a distorted version of the target signal $\sum_{i \in \mathcal{N}} \boldsymbol{s}_i^{k}$, owing to the AWGN and the partial user participation. 
The overall DPD-AirComp  architecture is shown in Fig.1, while the DPD-AirComp optimization procedure is summarized in Algorithm 1.
\begin{algorithm}[H]
\linespread{0.8}\selectfont
\begin{algorithmic}[1]\label{alg1}
\caption{DPD-AirComp} 
\State { {Initialize $\beta$, $\boldsymbol{x}_0$, $\boldsymbol{\lambda}_0$ and $a_0$}}
\For{ {$k=0,1,2,...K$}} 
    \State{The BS broadcasts the primal variable  {$\boldsymbol{x}^k$} to all users}
    \State{Users update the dual variable $\lambda^{k+1}_i$ based on (8)
    \State { {$\boldsymbol{s}_i^k \gets \lambda_i^k\boldsymbol{g}_i^k(\boldsymbol{x}^k)$} and  {$\boldsymbol{w}_i^k \gets \frac{\boldsymbol{s}_i^k}{\sqrt{\beta} h_i^k}, \quad \forall i \in \mathcal{N}$}}
    \If { {$\norm{\boldsymbol{w}_i^k}^2\leq P_\mathrm{max}$}}
      \State{The $i$-th user transmits $\boldsymbol{w}_i^k$} to the BS
    \Else
      \State{The $i$-th user does not participate}
    \EndIf
    \State{The BS receives  {$\tilde{\boldsymbol{y}}^k=\sum_{i \in \mathcal{A}^k}\boldsymbol{s}_i^{k}+\sqrt{\beta}\boldsymbol{n}^k$ }}
    \State{  {$\boldsymbol{x}^{k+1}\!\!=\!\! \mathcal{P}_\mathcal{X}\!\left[ \boldsymbol{x}^{k}\!\!-\!a_k\left( \boldsymbol{g}_0\left(\boldsymbol{x}\right) \!\!+\!\tilde{\boldsymbol{y}}^k\right) \right]$}}
    \State{ {$k\gets k+1$}}}
    \State{Update $a_k$}
\EndFor
\end{algorithmic}
\end{algorithm}
\section{Convergence Analysis} In this section, we examine the convergence behavior of the DPD-AirComp framework. First, we introduce some standard assumptions to facilitate the convergence analysis. 
\begin{assumption} 
For compact sets $\mathcal{X}$ and $\mathcal{D}$, and convex functions $f_0,f_1,...f_N$ over $\mathbb{R}^n$, the following uniform upper bound  exists \cite{nedic2},
\begin{equation}  
	\norm{\lambda_i\boldsymbol{g}_i(\boldsymbol{x}^k)}\leq G, \quad \forall i \in \mathcal{N}, \, \forall k.
\end{equation}
As a consequence, the subgradients of $\mathcal{L}$ are also uniformly bounded, i.e., there is a constant $L>G$ such that
\begin{equation}  
	\norm{\boldsymbol{\mathcal{L}}_x\left(\boldsymbol{x}^k,\boldsymbol{\lambda}^k\right)}\leq L \quad \text{and} \quad \norm{\boldsymbol{\mathcal{L}}_\lambda\left(\boldsymbol{x}^k,\boldsymbol{\lambda}^k\right)} \leq L, \quad \forall k.
\end{equation} 
\end{assumption}
\begin{assumption}
 Given that set $\mathcal{X}$ is compact, then for the primal optimal value $x^*$, and for any $k\geq 0$, a positive number $R$ exists, which satisfies $\mathbb{E}\left[\norm{\boldsymbol{x}^k-\boldsymbol{x}^{*}}\right]\leq R$. This assumption also implies that $\mathbb{E}\left[\boldsymbol{x}^k-\boldsymbol{x}^{*}\right]\preceq R$.
\end{assumption}
We note that Assumption 1 was proven in \cite{nedic2}, while Assumption 2 follows directly from the definition of compact sets. Moreover, we select a square summable, but not summable diminishing step size, i.e.,
\begin{equation}  
	a_k \geq 0, \quad \sum_{k=0}^{\infty}a_k^2 < \infty, \quad \sum_{k=0}^{\infty}a_k = \infty.
	\vspace{-.0cm}
\end{equation}
Next, let  the running weighted averages $\hat{\boldsymbol{x}}^k$ and $\hat{\boldsymbol{\lambda}}^k$, be
\begin{equation}  
	\begin{aligned}
		\hat{\boldsymbol{x}}^k=\frac{\sum\limits_{j=0}^{k-1}a_j\boldsymbol{x}^j}{\sum\limits_{j=0}^{k-1}a_j} \quad \text{and} \quad \hat{\boldsymbol{\lambda}}^k=\frac{\sum\limits_{j=0}^{k-1}a_j\boldsymbol{\lambda}^j}{\sum\limits_{j=0}^{k-1}a_j},
	\end{aligned}
	\vspace{-.0cm}
\end{equation}
which will be used to provide approximate solutions to problem (1), as well as convergence bounds for both the expected constraint violation and the optimality gap of the objective function. 
Based on these assumptions, we introduce the following theorems and lemmas:
\begin{theorem}
	The expected constraint violation of the proposed algorithm, at $\hat{\boldsymbol{x}}^k$, is bounded as
	\vspace{-.0cm}
    \begin{equation}   \label{conviol}
		\begin{aligned}
			&\mathbb{E}\left[ \norm{\left[\boldsymbol{F}\left(\hat{\boldsymbol{x}}^k\right)\right]^{+}}\right] \leq \frac{1}{rZ_k}\Bigg(RG\sum\limits_{j=0}^{k-1}\left(N-\bar{A}_j\left(\beta\right)\right)a_j \\\
            &+2\left(\frac{f_0\left(\bar{x}\right)\!\!-\!\!\tilde{q}}{\gamma} \!\!+\!\! r\right)^2 +\left(\beta\sigma^2+\frac{3L^2}{2}\right)\sum\limits_{j=0}^{k-1}a_j^2+\mathbb{E}\left[\norm{\boldsymbol{x}_0-\boldsymbol{x}^{*}}^2\right]\\
            &+2LG\sum\limits_{j=0}^{k-1}\left(N-\bar{A}_j\left(\beta\right)\right)a_j^2
			+L^2\sum\limits_{j=0}^{k-1}\left(N-\bar{A}_j\left(\beta\right)\right)^2a_j^2\Bigg),
		\end{aligned}
		\vspace{-.0cm}
	\end{equation}
	where $Z_k=\sum_{j=0}^{k-1}a_j$ and $\bar{A}_k(\beta)$ denotes the average number of participating users in the $k$-th round.
\end{theorem}
\begin{IEEEproof}
	See Appendix A.
\end{IEEEproof}

\begin{corollary}
 By selecting $r$, in the $k$-th round, as
\begin{equation}  
     r^{*}_k=\frac{\zeta+\sqrt{2\zeta^2+\delta_k Z_k}}{2},
\end{equation}
where $\zeta=\frac{f_0\left(\bar{x}\right)-\tilde{q}}{\gamma}$ and 
\begin{equation}    \label{eq:delta}
    \Scale[1]{
		\begin{aligned}
			\delta_k&= \frac{1}{Z_k}\Bigg(R G\sum\limits_{j=0}^{k-1}\left(N-\bar{A}_j\left(\beta\right)\right)a_j +\left(\beta\sigma^2+\frac{3L^2}{2}\right)\sum\limits_{j=0}^{k-1}a_j^2 \\
			 &+\mathbb{E}\left[\norm{\boldsymbol{x}_0-\boldsymbol{x}^{*}}^2\right] 
              +2LG\sum\limits_{j=0}^{k-1}\left(N-\bar{A}_j\left(\beta\right)\right)a_j^2 \\
              &+L^2\sum\limits_{j=0}^{k-1}\left(N-\bar{A}_j\left(\beta\right)\right)^2a_j^2\Bigg),
		\end{aligned}}
\end{equation}
we obtain
\begin{equation}  
\lim_{k \to \infty} \mathbb{E}\left[ \norm{\left[\boldsymbol{F}\left(\hat{\boldsymbol{x}}^k\right)\right]^{+}}\right] =0.
\end{equation}
\end{corollary}
\begin{IEEEproof}
	See Appendix B.
\end{IEEEproof}

\begin{theorem} An expected upper bound on the optimality gap, $f_0\left(\hat{\boldsymbol{x}}^k\right)-f_0^{*}$, where $f_0^{*}$ is the optimal value of the objective function, is given by
\begin{equation}   
\begin{aligned}\label{TH2}
&\mathbb{E}\left[f_0\left(\hat{\boldsymbol{x}}^k\right)-f^{*}_0\right]\leq \frac{RG}{Z_k}\Bigg(\frac{1}{2}\mathbb{E}\left[\norm{\boldsymbol{\lambda}^{0}}^2\right]+\mathbb{E}\left[\norm{\boldsymbol{x}_0-\boldsymbol{x}^{*}}^2\right]\\
&+\left(\beta\sigma^2+\frac{3}{2}L^2\right)\sum_{j=0}^{k-1}a_j^2+2LG\sum_{j=0}^{k-1}\left(N-\bar{A}_j\left(\beta\right)\right)a_j^2\\
 &+L^2\sum_{j=0}^{k-1}\left(N-\bar{A}_j\left(\beta\right)\right)^2a_j^2 \Bigg)
 \!\!+\!\!\frac{RG}{Z_k}\sum_{j=0}^{k-1}\left(N-\bar{A}_j\left(\beta\right)\right)a_j.
\end{aligned}
\end{equation}
\end{theorem}
\begin{IEEEproof}
	See Appendix C.
\end{IEEEproof}
\begin{theorem}
An expected lower bound on the optimality gap, $f_0\left(\hat{\boldsymbol{x}}^k\right)-f_0^{*}$, is given by
\begin{equation}  
\mathbb{E}\left[f_0\left(\hat{\boldsymbol{x}}^k\right)-f^{*}_0\right] \geq -\frac{f_0\left(\bar{x}\right)-q}{\gamma}\mathbb{E}\left[ \norm{\left[\boldsymbol{F}\left(\hat{\boldsymbol{x}}^k\right)\right]^{+}}\right].
\end{equation}
\end{theorem}
\begin{IEEEproof} The proof is similar to that of Theorem 1 and 2, and thus, omitted for brevity.
\end{IEEEproof}
\begin{remark}
According to Corollary 1, by optimally selecting the parameter $r$, the expected constraint violation tends to zero asymptotically. Therefore, the proposed AirComp implementation guarantees the feasibility of the obtained solution, i.e., the solution satisfies all the problem constraints. This is a notable result, indicating that the partial user participation in each round, stemming from the AirComp principles, does not affect the feasibility of the extracted solution.
\end{remark}
\begin{remark}
According to Theorem 3, the lower bound of the expected optimality gap depends on the value of the expected constraint violation $\mathbb{E}\left[ \norm{\left[\boldsymbol{F}\left(\hat{\boldsymbol{x}}^k\right)\right]^{+}}\right]$. Thus, by also considering Corollary 1, it is concluded that the lower bound of the expected optimality gap tends asymptotically to zero. Regarding the upper bound of the expected optimality gap in Theorem 2, it is easy to verify that only the last term on the right-hand-side (RHS) of \eqref{TH2} does not converge to zero, and thus, creates a non-zero optimality gap. Specifically, the considered term decreases as the average number of participating users increases and vanishes when $\bar{A}_j(\beta)=N$, i.e., all users participate in each round. In order to mitigate the effects of partial user participation and reduce the optimality gap, one should properly choose the value of the preprocessing scalar $\beta$. As can be observed in (17) and (60), higher values of $\beta$ force more users to participate in each round, on average. This policy increases the noise variance, i.e., $\beta\sigma^2,$ but the impact of noise vanishes as the number of rounds increases. To this end, the impact of $\beta$ on the convergence gap will be investigated numerically in Section V.
\end{remark}
\vspace{-0.45cm}
\section{Application of DPD-AirComp in Practical Use Cases}
In this section, we examine the application of the proposed DPD-AirComp algorithm in two distributed optimization use cases. The following analysis will shed light on the practical implementation of the proposed algorithm and serve as the basis for its performance evaluation in Section V. Specifically, we examine the following two application use cases:
\subsubsection{Use Case A} The energy management of a smart grid system, for which a Stackelberg game is formulated.
\subsubsection{Use Case B} The resource allocation for a conventional FDMA system, where the users' sum rate is to be maximized. 


\subsection{Smart Grid Energy Management}
\subsubsection{Model and problem formulation}
The considered system includes several primary and secondary load subscribers, as  well as a smart energy manager (SEM) \cite{grids}. After meeting the demands of the primary consumers, the smart grid wishes to sell its excess energy (if any) to the secondary subscribers connected to it, such as plug-in electric vehicles (PEVs). Set $\mathcal{N}$, consists of $N$ PEVs which can wirelessly communicate with the SEM. Also, we assume that the maximum amount of energy that the power grid can sell to the PEVs is $C$. The power grid aims  to optimize the distribution of its excess energy to the PEVs which maximizes its revenue. Hence, it will set an appropriate price $p$ per unit of energy. On the other hand, all PEVs, $\forall n \in \mathcal{N}$, intent to satisfy their energy demands, by requesting a certain amount of energy $u_n$ from the grid. That request varies among PEVs based on parameters like their battery capacity $b_n$. The interaction between the PEVs and the power grid can be modelled as a Stackelberg game. See  \cite{grids} for more details. The utility function of the grid is given by
\begin{equation}  
z\left(p,\boldsymbol{u}\right)=p\sum_{n=1}^Nu_n,
\end{equation}
and captures the total revenue of the grid when selling the energy required by all PEVs at a price $p$ per unit of energy. Vector $\boldsymbol{u}$ contains the energy demand $u_n$ of all $N$ subscribers. The SEM's goal is to maximize its utility function, which is equivalent to maximizing its profit. On the other side, for a fixed price $p$ the utility function of the $n$-th PEV is
\vspace{-.0cm}
\begin{equation}  
U_n\left(u_n\right) = b_nu_n-\frac{1}{2}s_nu_n^2-pu_n,
\end{equation}
where $s_n$ is a satisfaction parameter. The interaction between the SEM and the PEVs can be modelled as a two stage Stackelberg game, which is described as follows \cite{grids}

\vspace{-.0cm}
\emph{SEM stage:}
\begin{equation}  
\vspace{-.0cm}
\begin{aligned}
\underset{p}{\text{max}} \quad &  p\sum_{n=1}^Nu_n \\
\end{aligned}
\end{equation}

\emph{PEV stage (the following problem is solved $\forall n$):}
\begin{equation}   \label{pevgstage}
\vspace{-.0cm}
\begin{aligned}
\underset{u_n}{\text{max}} \quad &  b_nu_n-\frac{1}{2}s_nu_n^2-pu_n \quad \text{s.t.}  & \sum_{n=1}^Nu_n \leq C.
\end{aligned}
\end{equation}
According to \cite{grids}, problem (32) maximizes the revenue of the SEM, while the solution of problem (33), for the $n$-th PEV, yields its optimal energy demand $u_n$, with respect to the price per unit $p$. Given that all PEVs have obtained their optimal energy demand $u_n, \forall n \in N$, by solving problem \eqref{pevgstage}, then, the optimal price of the SG is given by \cite{grids} \\
\vspace{-.0cm}
\begin{equation}   \label{poweropt}
p^{*}=b_n-s_nu_n^{*}.
\vspace{-.0cm}
\end{equation} 
\par
Nonetheless, in large-scale smart grids having a significant number of PEVs, the communication overhead between the grid and the mobile consumers can be prohibitive. This is attributed to the global constraint in \eqref{pevgstage}, which implies that each PEV has to have knowledge of the energy demands $u_n, \forall n \in \mathcal{N}$, of all other PEVs. Thus, due to its scalability and low communication overhead, we will employ the proposed DPD-AirComp for solving problem (33). First we transform problem \eqref{pevgstage} to the form of (1). From \cite{grids}, the socially optimal Nash equilibrium of the PEV stage can  be found by solving the following problem
\begin{equation}   \label{pevgstage2}
\begin{aligned}
\underset{\boldsymbol{u}}{\text{max}} \quad   &\sum_{n=1}^NU_n \quad \\
\text{s.t.} \quad  &\sum_{n=1}^Nu_n \leq C.
\end{aligned}
\end{equation}
The objective value of (35) is the sum of the PEV's utility functions, while the constraint imposes a global constraint to the total energy demand of all PEVs.
Then, by rewriting problem \eqref{pevgstage2} into its epigraph form, we obtain
\begin{equation}   \label{pevgstageep}
\begin{aligned}
\underset{\boldsymbol{x}}{\text{max}} \quad & \sum_{n=1}^N y_n \quad \\
\text{s.t.} \quad &y_n  \leq U_n, \quad \\
&\boldsymbol{x} \in \mathcal{X}.
\end{aligned}
\end{equation}  
We define the vector  $\boldsymbol{y}$, which contains the $y_n$  for all $N$ users. By also defining $\boldsymbol{x} = \{\boldsymbol{u},\boldsymbol{y}\}$, $f_0(\boldsymbol{x})=-\sum_{n=1}^N y_n$, $f_n(\boldsymbol{x})= y_n-U_n, \forall n \in \mathcal{N}$, and the feasible set $\mathcal{X}$ of global constraints as follows  
\begin{equation*} 
\mathcal{X}=\left\{\boldsymbol{x} \, \Big| \, \sum_{n \in \mathcal{N}} u_n \leq C, \, \boldsymbol{u}\succeq \boldsymbol{0},\, \boldsymbol{y} \in \mathbb{R}^N \right\},
\end{equation*}
problem \eqref{pevgstageep} is convex and in the form of (1). Thus, its solution can be obtained according to Algorithm 1. 
After Algorithm 1 has converged, the power grid will obtain its optimal value $p^*$ according to \eqref{poweropt} and it will broadcast this value to all PEVs. Based on that new pricing, and by utilizing DPD-AirComp, all PEVs will renew their energy demands. This AirComp interaction between the SEM and the PEVs is repeated until an equilibrium between the power grid and the PEVs is reached. 

\subsubsection{Choice of set \texorpdfstring{$\mathcal{D}$}{D}}
It is noted that for many optimization problems finding the optimal set $\mathcal{D}$ as given in \eqref{setoptimal} may not be straightforward, or in some cases, as difficult as solving the original optimization problem itself. To address this challenge, we consider a suboptimal set $\mathcal{D}$. From the analysis in Section III.A, it can be observed that the optimal choice of $r^*_k$ holds for any $\zeta \geq 0$. Thus, in practice, instead of finding a point which satisfies the Slater conditions, an arbitrary point $\zeta{'} \geq 0$ can be selected. However, the optimal choice of $r^*_k$ still requires  knowledge of parameters which may be unknown in practice, such as the bound $L$ of the subgradients. However, from Section III.A and \eqref{conviol}, it can be verified that a choice of the form
\begin{equation}  
r_k=\vartheta \cdot \sqrt{\sum_{j=0}^{k-1}a_j}, \quad \vartheta>0,
\end{equation} guarantees that $\lim_{k \to \infty} \mathbb{E}\left[ \norm{\left[\boldsymbol{F}\left(\hat{x}_k\right)\right]^{+}}\right] =0$. Hence, in practice, the following  set can be used
\vspace{-.0cm}
\begin{equation}   \label{choiceofD}
\mathcal{D}_k=\left\{\lambda \geq 0 \bigg| \norm{\lambda}_{\infty} \leq \zeta{'} + \vartheta \sqrt{\sum_{j=0}^{k-1}a_j}\right\},
\vspace{-.0cm}
\end{equation}
while the impact of the constant values $\zeta{'}$ and $\vartheta$ on the convergence will be studied numerically. Also, the value of the preprocessing factor $\beta$ will be determined based on a few Monte Carlo iterations. Specifically,  the value of $\beta$ which provides the best average DPD-AirComp  performance will be selected, while its impact on the obtained solution will be numerically investigated.

\subsubsection{Projection onto sets \texorpdfstring{$\mathcal{D}$}{D} and \texorpdfstring{$\mathcal{X}$}{X}}
In this subsection, the projection onto sets $\texorpdfstring{\mathcal{D}}{D}$ and $\texorpdfstring{\mathcal{X}}{X}$ will be discussed. The projection of $\boldsymbol{\lambda}$ onto set $\texorpdfstring{\mathcal{D}}{D}$ is straightforward. On the contrary, the projection of $\boldsymbol{x}$ is not straightforward. 
Set $\texorpdfstring{\mathcal{X}}{X}$ imposes two sets of inequality constraints on vector $\boldsymbol{u}$, which do not necessarily hold with  equality, and thus, there is no closed-form expression to describe the projection explicitly. Still, it can be computed by standard convex optimization tools, such as second-order methods, at the expense of an increased complexity, roughly of $\texorpdfstring{\mathcal{O}}{O}\left(N^3\right)$ for interior-point methods \cite{intpoint}. Therefore, we propose an alternative low-complexity projection algorithm for vector $\boldsymbol{u}$. The projection process in the $k$-th round can be formulated as follows
\begin{equation}   \label{eq:proj}
\begin{aligned}
\underset{\boldsymbol{u}}{\text{min}} \quad & \norm{\boldsymbol{u}-\boldsymbol{u}^{k}}^2 \quad\\
\text{s.t.} \quad &\sum_{n=1}^Nu_n \leq C, \quad \\
&\boldsymbol{u} \succeq 0. 
\end{aligned}
\end{equation}
To solve \eqref{eq:proj}, first we project $\boldsymbol{u}^k$ onto set $\mathbb{R}^N_+$, stemming from the constraint $\boldsymbol{u} \succeq \boldsymbol{0}$. As a consequence, we obtain $\boldsymbol{u}_\mathrm{new}^{k}=\left[\boldsymbol{u}^{k}\right]^+$, and compute $\sum_{n=1}^Nu_{n,\mathrm{new}}^{k}=C{'}$. If $C{'}\leq C$, then the projection process terminates, since a feasible solution has been found. In the following, we assume that $C{'} > C$. Also, we define set 
\begin{equation}  
    \mathcal{M}=\left\{n \in \mathcal{N} \,  \Big\vert \, u^k_{n,\mathrm{new}}=0 \right\}\,\, \text{and} \,\,  \mathcal{N{'}}=\mathcal{N} \setminus \mathcal{M}.
    \vspace{-.0cm}
\end{equation} 
Then, the projection problem can be rewritten as 
\begin{equation}   \label{eq:proj2}
\begin{aligned}
\underset{\boldsymbol{u}\in \mathcal{N{'}}}{\text{min}} \quad & \norm{\boldsymbol{u}-\boldsymbol{u}_\mathrm{new}^{k}}^2 \quad \\
\text{s.t.} &\quad \sum_{n=1}^{\lvert \mathcal{N{'}}\rvert}u_n \leq C, \\
\end{aligned}
\end{equation}
where, hereinafter, $\boldsymbol{u}_\mathrm{new}^{k}$ contains only the terms for which $u_{n,\mathrm{new}}^{k}>0, \forall n \in \mathcal{N}^{'}$, holds. We note that problem (41) is convex. By introducing a Lagrange multiplier $\psi$ for the inequality constraint and applying the KKT conditions, it is straightforward to prove that at the optimal point of (41) it holds that 
\begin{equation}   \label{eq:kkt}
\begin{aligned}
&u_n^{*}=\frac{2u_\mathrm{new,n}^{k}-\psi^*}{2}\,\,, \forall n \in N,\,\,\text{and}\\
\quad &\psi^{*}\left(\sum_{n=1}^{N{'}}u_n^{*} - C\right)=0.
\end{aligned}
\end{equation}
According to the complementary slackness condition, there are two possible outcomes that need to be investigated. The first is
\begin{equation}  
\begin{aligned}
&\psi^{*}=0 \quad \text{and} \quad \sum_{n=1}^{\lvert \mathcal{N{'}}\rvert}u_n^{*} - C < 0, 
\end{aligned}
\end{equation}
which according to \eqref{eq:kkt} further implies that $\sum_{n=1}^N u_{\mathrm{new},n}^{k}<C$, and therefore, the initial vector of $\boldsymbol{u}_\mathrm{new}^{k}$ is already in set $\mathcal{U}$, which is false by contradiction. As such, the following has to hold 
\begin{equation}  
\begin{aligned}
&\psi^{*}>0 \quad \text{and} \quad \sum_{n=1}^{\lvert \mathcal{N{'}}\rvert}u_n^{*} - C = 0. 
\end{aligned}
\end{equation} 
From the equality condition and relation \eqref{eq:kkt} we have
\begin{equation}  
\sum_{n=1}^{N{'}} u_n^{*}=C \Leftrightarrow \psi^{*}=\frac{2}{\lvert \mathcal{N{'}}\rvert}\left(C{'}-C\right),
\end{equation}
and from the optimal $\psi^{*}$ the projected version of $\boldsymbol{u}_\mathrm{new}^{k}$ is given below by
\begin{equation}  
u_n^{*}=u_{\mathrm{new},n}^{k}-\frac{C{'}-C}{\lvert \mathcal{N{'}}\rvert}, \forall n \in \mathcal{N}.
\end{equation}
The overall projection process is summarized in Algorithm 2. The worst case complexity of Algorithm 2 can be found equal to $\mathcal{O}\left(N^2\right)$.
\begin{algorithm}[H]
\linespread{0.8}\selectfont
\caption{Proposed Projection Algorithm}
\begin{algorithmic}[1]\label{alg2}
\State{ {Given $\boldsymbol{u}^{k}, C$}}
\State{ {$\boldsymbol{u} \gets \left[\boldsymbol{u}^{k}\right]^+$}}
\State{ {$\mathcal{N{'}}=\mathcal{N}$}}
\While { { $\exists n \in \mathcal{N} \mkern9mu \text{for which} \mkern9mu u_n<0 \mkern4mu$ or $\sum_{n=1}^N u_n > C$}}
    \State{ {$ \boldsymbol{u} \gets \left[\boldsymbol{u}\right]^+$}}
    \State{ {$ \mathcal{N{'}}=\{u_n \vert  u_n>0, \forall n \in \mathcal{N} \}$}}
    \State { {$\sum_{n=1}^{\lvert\mathcal{N{'}}\rvert} u_n=C{'}$}}
    \State{ {$u_n \gets u_n-\frac{C{'}-C}{\lvert \mathcal{N{'}}\rvert}\,\,, \forall n \in N $}}
\EndWhile
\State{ {$\boldsymbol{u}^{*} \gets \boldsymbol{u}$}}
\end{algorithmic}
\end{algorithm}

\subsection{FDMA Joint Power and Bandwidth Allocation}
As a second use case, we consider a classic wireless communication resource allocation problem. Specifically, we focus on maximizing the sum rate of network users, in a distributed manner, by jointly optimizing the power and bandwidth allocation. We assume $N$ mobile users, and a total of $K$ frequency bands. The power, the fraction of the bandwidth and the channel coefficient of the $n$-th user in the $k$-th frequency band are denoted as $p_{n,k}$, $w_{n,k}$, and $h_{n,k}$, respectively. The considered problem is formulated as follows
\begin{equation}   \label{ofdmconvex}
\begin{aligned}
\underset{\boldsymbol{p},\boldsymbol{w}}{\text{max}} \quad &  \sum_{k=1}^K\sum_{n=1}^N R_{k,n} \quad\\
&\text{s.t.} \quad  R_n\geq R_\mathrm{th},\\
&\sum_{k=1}^K\sum_{n=1}^Np_{k,n} \leq P, \,\\
&\sum_{k=1}^K w_{k,n} = 1,\,\\
&p_{k,n}\geq 0,\, w_{k,n} \geq 0,
\end{aligned}
\end{equation}
where $R_{k,n}=\frac{w_{k,n}B}{K}\left(1+\frac{p_{n,k}h^2_{n,k}}{w_{n,k}N_0\frac{B}{K}}\right)$, $R_n=\sum_{n=1}^NR_{k,n}$, and $R_\mathrm{th}$ is a quality of service threshold. Also,  $\mathbf{p}$ and $\mathbf{w}$ denote the vectors which contain all variables $p_{k,n}$ and $w_{k,n}$, respectively, $N_0$ denotes the power spectral density of the AWGN,  and $B$ denotes the total bandwidth available at the BS. It is easy to verify that the problem is convex. Moreover, we define $\boldsymbol{x} = \{\boldsymbol{p},\boldsymbol{w}\}$, $f_0(\boldsymbol{x})=\!\!-\sum_{k=1}^K\sum_{n=1}^N R_{k,n}$, $f_n(\boldsymbol{x})=R_\mathrm{th}-R_n, \forall n \in \mathcal{N}$, and the feasible set $\mathcal{X}$ of global constraints as follows  
\begin{equation*} 
\mathcal{X}=\left\{\boldsymbol{x} \, \Bigg| \, \sum_{k=1}^K\sum_{n=1}^Np_{k,n} \leq P,\, \sum_{k=1}^K w_{k,n} = 1 , \, \boldsymbol{p}\succeq \boldsymbol{0},\, \boldsymbol{w}\succeq \boldsymbol{0} \right\}.
\end{equation*}
We note that problem (47) is in the form of problem (1), thus, based on Algorithms 1 and 2, its DPD-AirComp implementation is straightforward.

\begin{table}
    \linespread{0.9}\selectfont
    \centering
    \caption{Simulation Parameters}
    \label{table}
    \begin{tabular}{ |P{2cm}|P{2cm}|P{2cm}| }\hline
    \textbf{Parameter} & \textbf{Case A} & \textbf{Case B}\\ \hline\hline
    $B$ & 1MHz &   1MHz  \\  \hline
    $\sigma^2$ & -90dBm &  -90dBm \\ \hline
    $N$   & 20    & 10   \\ \hline
    $P_\mathrm{max}$ & 1W &  1W \\ \hline
    $\zeta$ & 2 &  2 \\ \hline
    $\theta$ & 2 &  1 \\ \hline
    $\beta$ & $10^4$ & $10^6$ \\ \hline
    $d/d_0$ & $[10,20]$ & $[10,20]$ \\ \hline
    $a$ & 2.2 & 2.2 \\ \hline
    $\epsilon$ & 10 & 10 \\ \hline
    $T_0$ & -25dB & -25dB \\ \hline
    $b_n$ & $[35,65]$MWh & -\\  \hline
    $s_n$   &$[1,2]$ & -\\  \hline
    $C$  & 99MWh  & -\\  \hline
    $P$ & - & 1W \\ \hline 
    $N_0$ & -  & $-174$dBm/Hz  \\  \hline
    $K$&   -      & 64   \\ \hline
    $R_{\mathrm{th}}$ & - &  2.85Mbps \\ \hline
\end{tabular}
\end{table}
\section{Performance Evaluation and Discussion}
In this section,  simulation results are provided to evaluate the performance of DPD-AirComp for both use cases A and B in Section IV. Also, we compare the  performance of the proposed scheme with a benchmark employing error-free transmission, as defined below.

\textit{Error-free transmission:} A digital orthogonal multiple access communication scheme is considered, where all users transmit with a rate below the Shannon capacity limit, with transmit power $P_\mathrm{max}$. Without loss of generality, we assume the TDMA communication protocol.

For the links between all devices and the BS, according to \cite{rice}, we assume Rician fading, $ 
\boldsymbol{h}\!\!\!\!= \!\!\!\!\sqrt{T_0(d/d_0)^{-a}}\left(\sqrt{\frac{\epsilon}{\epsilon + 1}}\boldsymbol{h}^{\mathrm{LoS}}+\sqrt{\frac{1}{\epsilon + 1}}\boldsymbol{h}^{\mathrm{NLoS}}\right)$, with Rician factor $\epsilon$, where $\boldsymbol{h}^{\mathrm{LoS}}$ and $\boldsymbol{h}^{\mathrm{NLoS}}$ denote the line-of-sight  and the non-line-of-sight components, respectively. Moreover, $T_0$ is the path loss at the reference distance of $d_0$, $d$ denotes the distance between the transmitter and the receiver, and $a$ is the path loss exponent. Unless specified otherwise, the selected learning step for use case A is $a_k=\frac{2}{3+k}$, and it is $a_k=\frac{1}{10^5+k}$ for use case B, while all simulation parametes are given in Table I.
All results have been averaged over 500 individual runs in a Monte Carlo fashion. Both choices satisfy the square summable but not summable diminishing step size requirement, as in (22). The initial points of Algorithm 1 are chosen randomly. 


Choosing an appropriate step size for subgradient methods is of paramount importance. Despite their theoretical convergence properties, in practice subgradient methods often do not achieve an exact zero constraint violation gap for all step size choices \cite{boyd2003subgradient}.
In Fig. 2b, the constraint violation for different step size choices is illustrated, for use case B. It is observed that for bigger step sizes the algorithm converges more rapidly to a stationary point, nonetheless, the achieved performance is poor. In constrast, for step size $a_k=\frac{1}{10^5+k}$, the constraint violation of DPD-AirComp converges to zero in 50 iterations, which is also the case for the error-free case.
Moreover, in Fig. 2a, the maximized sum rate for use case B is illustrated for the same step size choices, as considered in Fig. 2b. It can be seen that, for all considered step sizes, DPD-AirComp achieves an inferior sum rate compared to the identical error-free scheme, while for $a_k=\frac{1}{10^5+k}$ it closely approaches the error-free scheme. It is noted that the optimality gap is attributed to partial user participation, since according to the combination of Theorems 2 and 3 the optimality gap becomes equal to zero when all users participate in the optimization procedure.

\begin{figure}[t!]
\vspace{-0cm}
\centering
\begin{subfigure}[b]{.7\linewidth}
\includegraphics[width=1\linewidth]{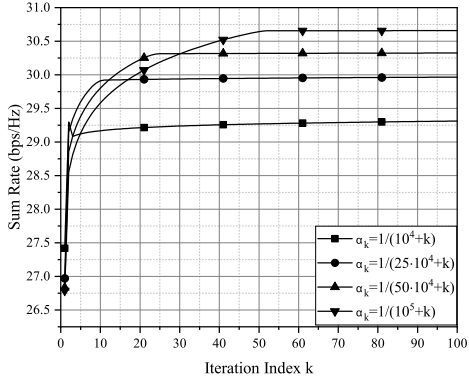}
\caption{Sum rate maximization.\vspace{0cm}}
\label{fig:fig3}
\end{subfigure}
\begin{subfigure}[b]{.7\linewidth}
\includegraphics[width=1\linewidth]{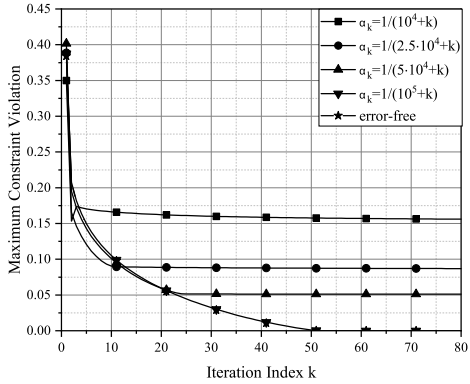}
\caption{Maximum constraint violation.} \label{fig:fig4}
\end{subfigure}
\vspace{-0cm}
\caption{The impact of step size $a_k$ for use case B.}
\vspace{-0cm}
\end{figure}

\begin{figure}[t!]
\centering
\begin{subfigure}[h]{.7\linewidth}
\centering
\includegraphics[width=1\linewidth]{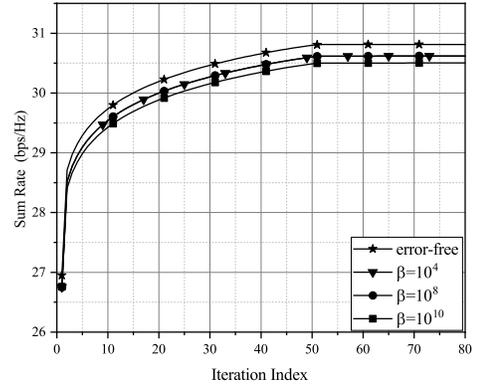}
\caption{Expected optimality gap.}
\label{fig:fig1}
\end{subfigure}

\begin{subfigure}[h]{.7\linewidth}
\centering
\includegraphics[width=1\linewidth]{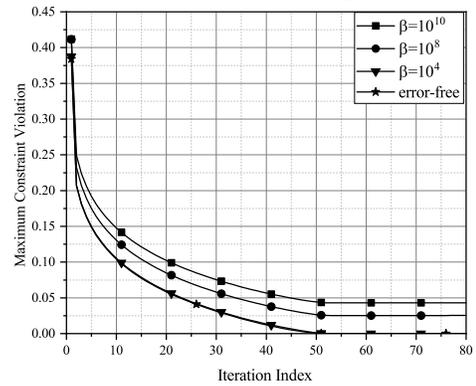}
\caption{Maximum constraint violation.} \label{fig:fig2}
\end{subfigure}
\vspace{-0in}
\caption{The impact of preprocessing factor, $\beta$, for use case B.}
\vspace{-0cm}
\end{figure}

In Fig. 3, the impact of the preprocessing factor $\beta$ is investigated for use case B. As can be seen, the value of $\beta$ affects both the expected maximum constraint violation of the problem and the optimal value. According to the convergence analysis in Section III, a larger value of $\beta$ enables more users to participate in each round, however, it also increases the effective noise power at the BS. For $\beta=10^4$ and $\beta=10^8$, the value of the objective function approximates the one of the error-free scheme, while for $\beta=10^{10}$, the optimality gap is larger, since $\beta=10^{10}$ increases the noise power levels. That is also the case in Fig. 3b, where it is observed that the expected maximum constraint violation for $\beta=10^{10}$ is approximately double compared to the case of $\beta=10^8$. For $\beta=10^4$ the constraint violation goes to zero. Therefore, the trade-off between the number of participating users per round and the noise power level at the receiver should be carefully balanced, when the value of $\beta$ is selected.

The considered trade-off is further showcased in Fig. 4, where the impact of $\beta$ for use case A is studied. Specifically, from Fig. 4a, it is evident that by selecting $\beta=10^6$, the DPD-AirComp algorithm achieves identical expected constraint violation as the error-free scheme, while asymptotically achieving a zero constraint violation. However, that is not the case for the other values of $\beta$. For instance, for $\beta=10^4$, the AirComp scheme does not converge, since only a few users participate in each round. Furthermore, for $\beta=10^8$, the AirComp scheme performs worse than for $\beta=10^6$. For $\beta=10^8$, although an increased number of users participate in each round, the increased noise level slows down the convergence of the algorithm and causes a larger constraint violation.  We note that  Theorem 1 states that for a proper selection of $a_k,r_k$, and $\beta$ the solution obtained from DPD-AirComp asymptotically achieves a zero expected constraint violation. From Fig. 3b and Fig. 4a this can be seen to hold for both investigated use cases. 

In Fig. 4b, the price per MWh obtained from the Stackelberg scheme is plotted for various values of $\beta$. It is noticed that for $\beta=10^6$, the Stackelberg price obtained with DPD-AirComp is equal to the error-free scheme. On the other hand, for $\beta=10^8$, the price value is less than the optimal price value of the error-free scheme, and eventually the power grid looses revenue. Nonetheless, this is a favorable condition for the PEVs, since they can purchase energy cheaper. This can be attributed to the fact that the optimal price of $p$, according to (34), depends on the solution of problem (36). However, due to AWGN and channel fading, the solution obtained by DPD-AirComp can be suboptimal, thus, resulting in a suboptimal price $p$.

\begin{figure}[t]
\vspace{-1cm}
\centering
\begin{subfigure}[t]{.7\linewidth}
\includegraphics[width=1\linewidth]{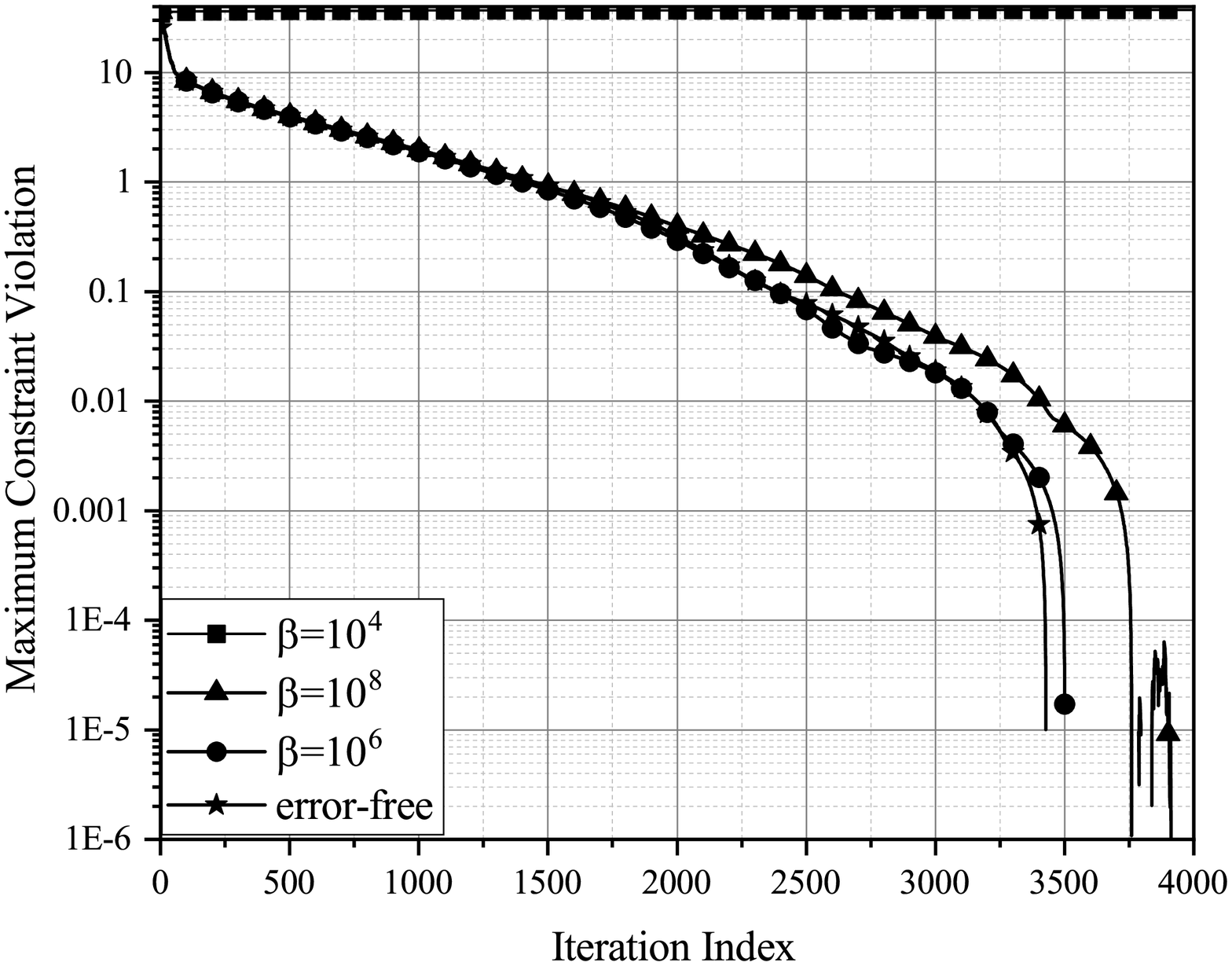}
\caption{Maximum constraint violation.}
\label{fig:fig5}
\end{subfigure}

\centering
\begin{subfigure}[t]{.7\linewidth}
\includegraphics[width=1\linewidth]{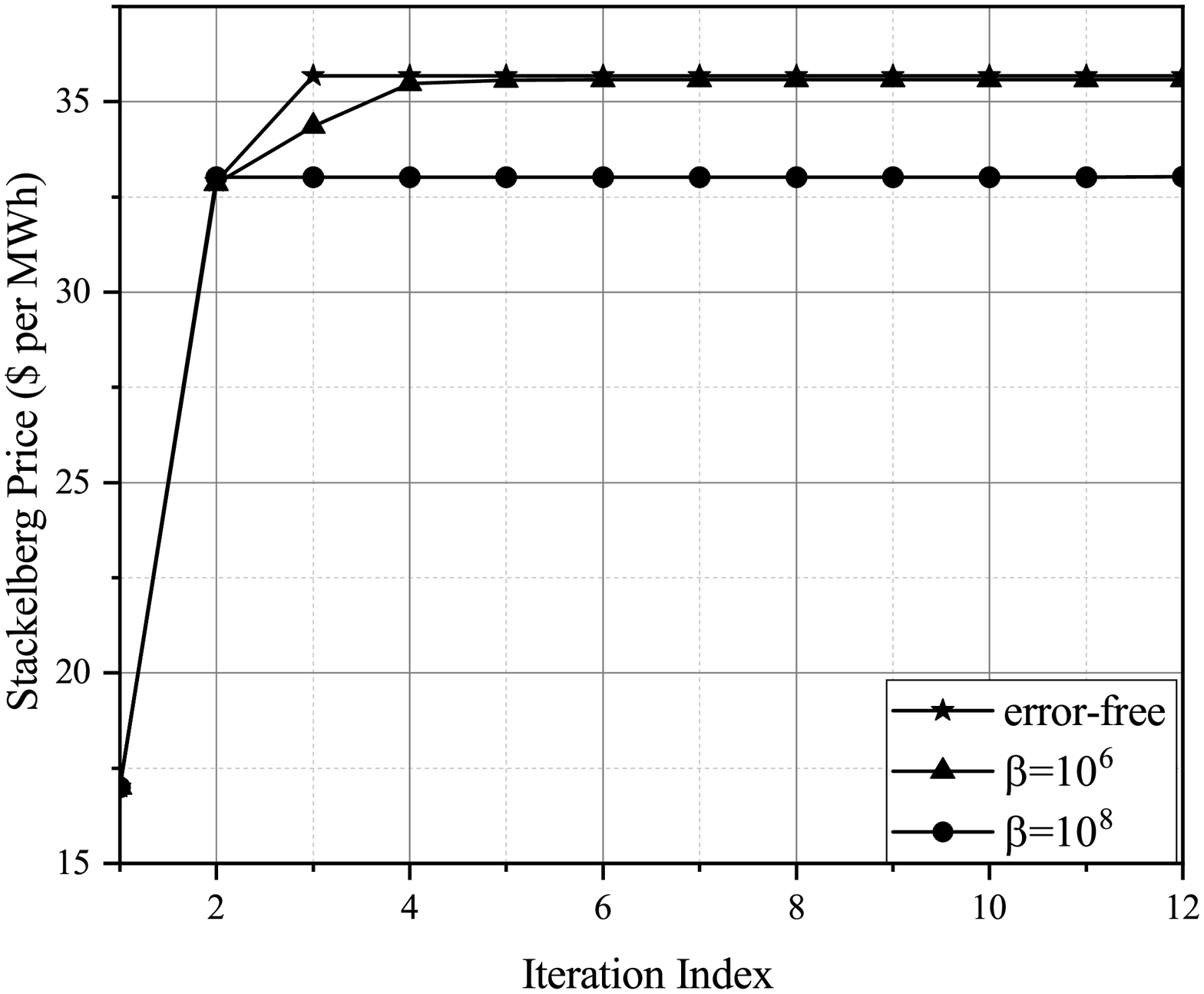}
\caption{Stackelberg price optimization.} \label{fig:fig6}
\end{subfigure}
\vspace{-0cm}
\caption{The impact of preprocessing factor, $\beta$, for use case A.}
\end{figure}

Moreover, in Fig. 5 the impact of set $\mathcal{D}$, as given in \eqref{choiceofD},  is illustrated. Set $\mathcal{D}$ bounds the maximum value of $\boldsymbol{\lambda}$. The values of $\boldsymbol{\lambda}$ are used to regulate the constraint violation of the problem, and Theorem 1 shows that the choice of $\mathcal{D}$ influences the expected constraint violation. Indeed, from Fig. 5 it can be verified that the choice of set  $\mathcal{D}$ is of paramount importance. For $\zeta=1,\theta=1$ it is seen that DPD-AirComp does not converge at all, while for $\zeta=3,\theta=3$ and $\zeta=5,\theta=5$, DPD-AirComp is not stable, eventually, converging to a constraint violation of $10^{-3}$. On the other hand, for $\zeta=2,\theta=2$, the constraint violation of the DPD-AirComp algorithm closely approaches that of the error-free scheme.

In Fig. 6, we compare the convergence rate, w.r.t. the unit of time in seconds, between the proposed DPD-AirComp scheme and the error-free baseline. For DPD-AirComp scheme, time duration of one communication round between the BS and all users, is given as $\frac{L}{B}$, where $L$ is the total number of symbols to be transmitted. For use case B, $L=2K$. The time duration of one iteration for the error-free scheme is calculated as 
$d_s=\sum_{i=1}^N \frac{64+L(1+\log_2(1+q_1))}{\log_2(1+\frac{P_\mathrm{max}|h_i|^2}{N_0B})},$
where $ \log_2(1+q_1)=16$, representing the quantization level \cite{amiri2020federated}. As can be seen, the DPD-AirComp scheme significantly outperforms the error-free scheme, in terms of convergence rate. Specifically, DPD-AirComp converges within the first 0.05 seconds, while the error-free baseline converges after 1 second. Therefore, the DPD-AirComp scheme is an order of magnitude faster than the error-free scheme, while achieving a near-optimal performance. 

\begin{figure}[t!]
\vspace{-1cm}
\centering
\begin{minipage}{.7\columnwidth}
  \includegraphics[width=1\linewidth]{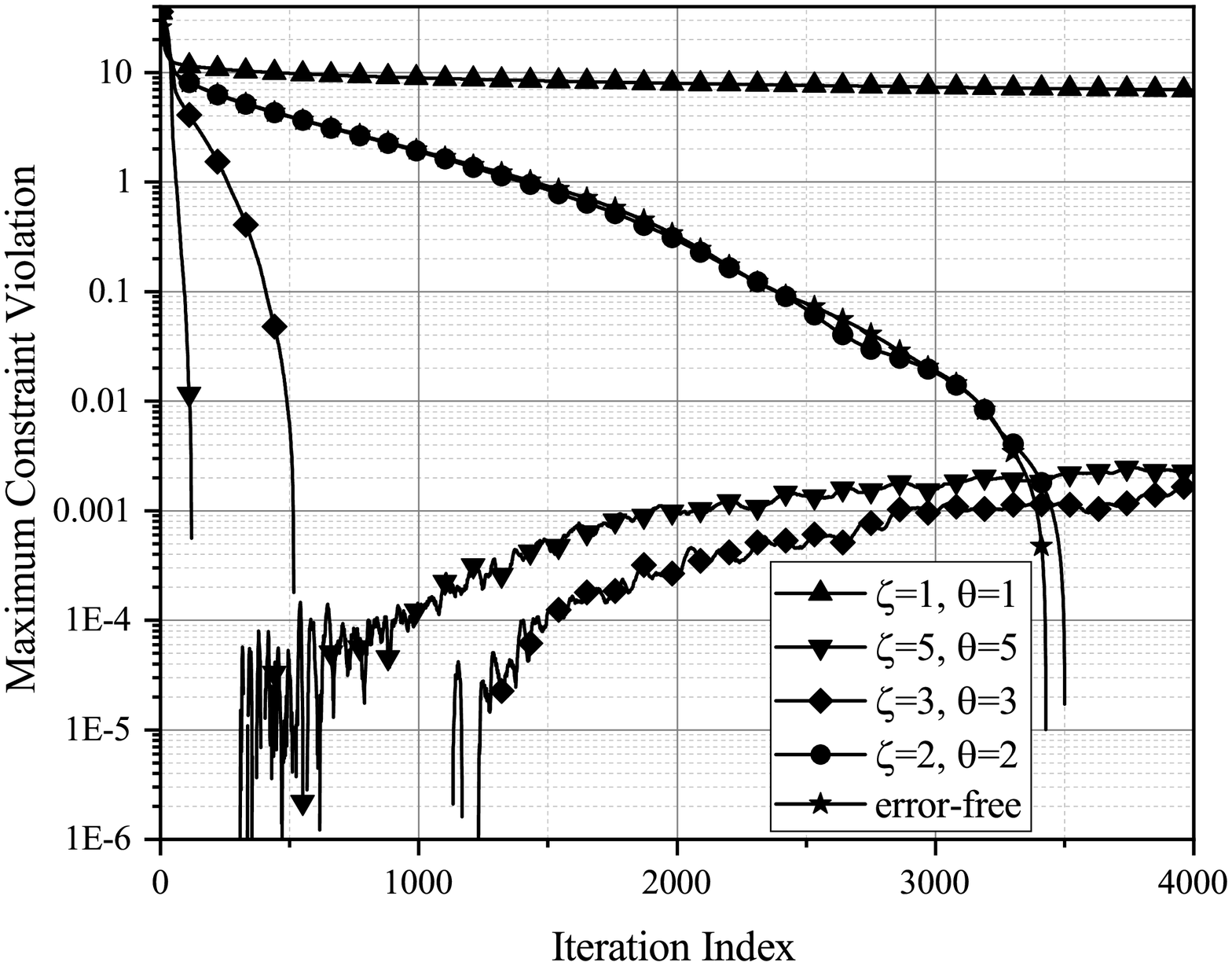}
  \caption{\vspace{-0in}Impact~of~set~$\mathcal{D}$~on~the~expected~constraint violation.}
  \label{fig:test1}
\end{minipage}%

\begin{minipage}{.7\columnwidth}
  \centering
  \includegraphics[width=1\linewidth]{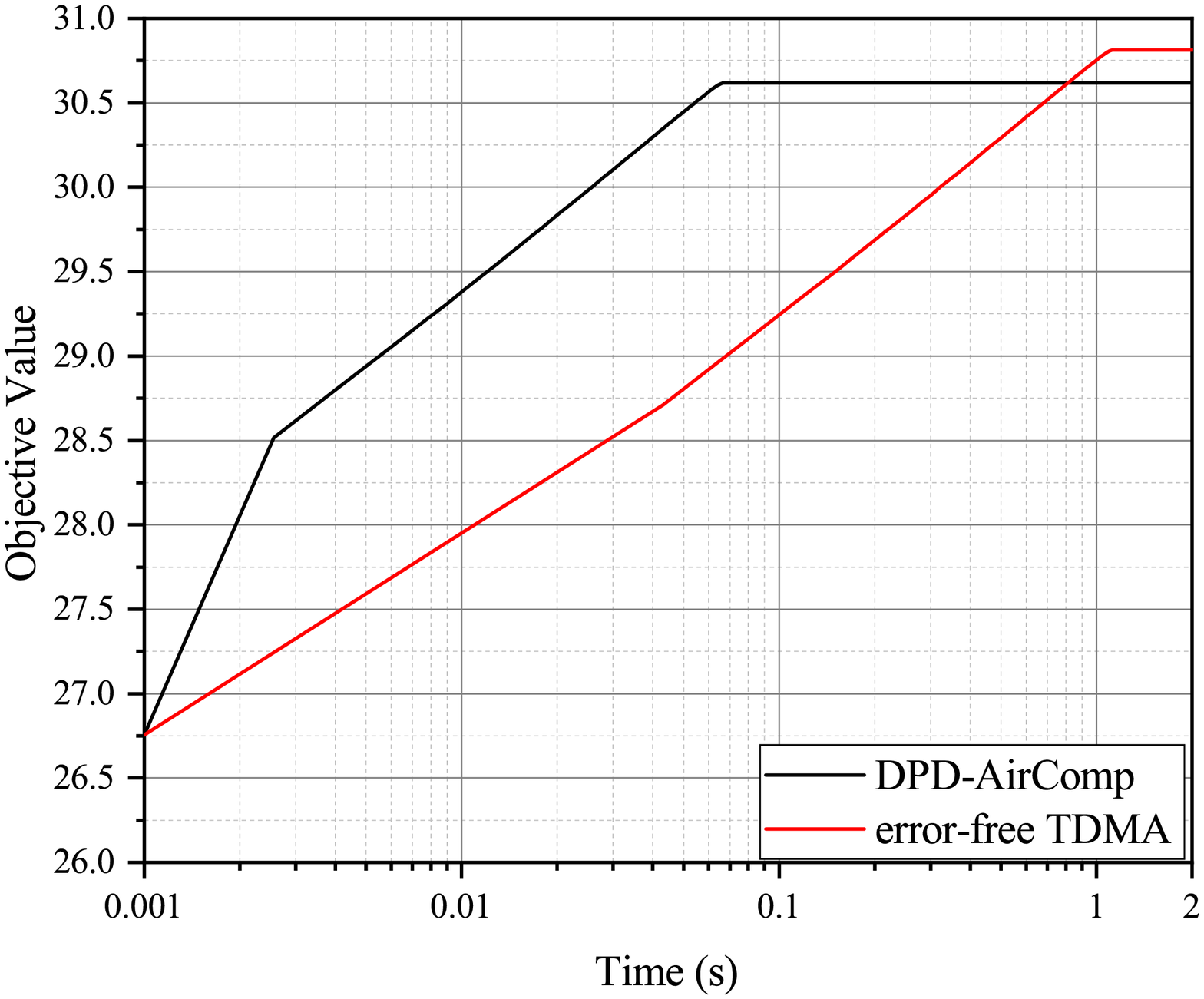}
  \caption{\vspace{-0in}Convergence time comparison.}
  \label{fig:test2}
\end{minipage}
\vspace{-0cm}
\end{figure}
\section{Conclusion}
In this paper, we proposed the DPD-AirComp framework for distributed optimization over the wireless medium. Assuming convexity, but not necessarily differentiability, and for a general objective function, we proved that the proposed DPD-AirComp can asymptotically achieve zero expected constraint violation. Therefore, DPD-AirComp ensures the feasibility of the original problem, despite the presence of channel fading and additive noise. 
Moreover, with proper power control of the users' signals, the non-zero expected optimality gap can be reduced. 
Furthermore, two practical use cases were presented, namely, smart grid energy management and FDMA resource allocation, and the implementation of DPD-AirComp for those use cases was extensively discussed. Finally, simulation results were presented which evaluate the performance of the DPD-AirComp scheme. Specifically, it was shown that the errors caused due to fading and noise can be properly handled by DPD-AirComp, while its convergence time is an order of magnitude faster compared to that of an error-free TDMA scheme.
\vspace{-.5cm}
\section*{Appendix A \\ Proof of Theorem 1}
We begin the proof by first proving some key lemmas.

\emph{Lemma 1a:} For all $\boldsymbol{x} \in \mathcal{X}$ and $k \geq 0$, we have
\begin{equation}   \label{eq:lemma1a}
	\begin{aligned}
		&\mathbb{E}\left[\norm{\boldsymbol{x}^{k+1}-\boldsymbol{x}^{*}}^2 \right] \leq \!\!
		-2a_k\mathbb{E}\left[\mathcal{L}\left(\boldsymbol{x}^k,\boldsymbol{\lambda}^k\right)-\mathcal{L}\left(\boldsymbol{x}^{*},\boldsymbol{\lambda}^k\right)\right]\!\\
        &+ \!\!\mathbb{E}\left[\norm{\boldsymbol{x}^{k}\!-\!\boldsymbol{x}^{*}}^2\right]\! +\!a_k^2\left(2LG\left(N-\bar{A}_k\left(\beta\right)\right)\!+\!\beta\sigma^2\!+\!L^2\right)\!\! \\
  &+2a_k\left(N-\bar{A}_k\left(\beta\right)\right)RG +a_k^2L^2\left(N-\bar{A}_k\left(\beta\right)\right)^2.
	\end{aligned}
\end{equation}
\begin{IEEEproof}
 We take
 \begin{equation}  
 	\begin{aligned}\label{x-xopt}
 		\norm{\boldsymbol{x}^{k+1}-\boldsymbol{x}^{*}}^2 &= \norm{ \mathcal{P}_\mathcal{X}\left[\boldsymbol{x}^k-a_k\tilde{\boldsymbol{\mathcal{L}}}_x\left(\boldsymbol{x}^k,\boldsymbol{\lambda}^k\right)\right]-\boldsymbol{x}^{*}}^2\\
   &\leq \norm{\boldsymbol{x}^k-a_k\tilde{\boldsymbol{\mathcal{L}}}_x\left(\boldsymbol{x}^k,\boldsymbol{\lambda}^k\right)-\boldsymbol{x}^{*}}^2,
 	\end{aligned}	
\end{equation}
where $\tilde{\boldsymbol{\mathcal{L}}}_x$ is given by
\begin{equation}  
	\begin{aligned}
		&\tilde{\boldsymbol{\mathcal{L}}}_x\left(\boldsymbol{x}^k,\boldsymbol{\lambda}^k\right)=\boldsymbol{g}_0\left(\boldsymbol{x}^k\right)+\sum_{i\in \mathcal{A}^k} \lambda_i \boldsymbol{g}_i\left(\boldsymbol{x}^k\right)+\sqrt{\beta}\boldsymbol{n}^k \\
		&=\boldsymbol{\mathcal{L}}_x\left(\boldsymbol{x}^k,\boldsymbol{\lambda}^k\right)- \sum_{i\in \mathcal{N} \setminus \mathcal{A}^k} \lambda_i \boldsymbol{g}_i\left(\boldsymbol{x}^k\right) +\sqrt{\beta}\boldsymbol{n}^k \\ &= \boldsymbol{\mathcal{L}}_x\left(\boldsymbol{x}^k,\boldsymbol{\lambda}^k\right) + \hat{\boldsymbol{\mathcal{L}}}_x\left(\boldsymbol{x}^k,\boldsymbol{\lambda}^k\right),
	\end{aligned}
\end{equation}
and
\begin{equation}  
	\hat{\boldsymbol{\mathcal{L}}}_x\left(\boldsymbol{x}^k,\boldsymbol{\lambda}^k\right)=-\sum_{i\in \mathcal{N} \setminus \mathcal{A}^k} \lambda_i \boldsymbol{g}_i\left(\boldsymbol{x}^k\right) +\sqrt{\beta}\boldsymbol{n}^k.
\end{equation}
Therefore, \eqref{x-xopt} leads to
\begin{equation}   \label{eq:p}
	\begin{aligned}
		\norm{\boldsymbol{x}^{k+1}\!-\!\boldsymbol{x}^{*}}^2  \!&\leq\! \norm{\boldsymbol{x}^{k}\!-\!\boldsymbol{x}^{*}}^2 \!-\!2a_k\underbrace{\inprod{\boldsymbol{\mathcal{L}}_x\left(\boldsymbol{x}^k,\boldsymbol{\lambda}^k\right),\boldsymbol{x}^k\!-\!\boldsymbol{x}^{*}}}_{I_1}\\
		&-2a_k\underbrace{\inprod{\hat{\boldsymbol{\mathcal{L}}}_x\left(\boldsymbol{x}^k,\boldsymbol{\lambda}^k\right),\boldsymbol{x}^k-\boldsymbol{x}^{*}}}_{I_2} \\ &+a_k^2\underbrace{\norm{\boldsymbol{\mathcal{L}}_x\left(\boldsymbol{x}^k,\boldsymbol{\lambda}^k\right)+\hat{\boldsymbol{\mathcal{L}}}_x\left(\boldsymbol{x}^k,\boldsymbol{\lambda}^k\right)}^2}_{I_3}.
	\end{aligned}
\end{equation}
By the definition of the subgradient in (6), the term $I_1$ on the RHS of \eqref{eq:p} is bounded as
\vspace{0cm}
\begin{equation}  \label{I1}
	\begin{split}
		&I_1
		\leq -\mathcal{L}\left(\boldsymbol{x}^k,\boldsymbol{\lambda}^k\right)-\mathcal{L}\left(\boldsymbol{x}^{*},\boldsymbol{\lambda}^k\right).
	\end{split}
\end{equation}
Moreover, $I_2$ can be bounded as
\begin{equation}  
	\begin{aligned}\label{I2}
		I_2\!&=\!-\!\!\!\sum_{i\in \mathcal{N} \setminus \mathcal{A}^k} \inprod{\lambda_i\boldsymbol{g}_i\left(\boldsymbol{x}^k\right),\boldsymbol{x}^{k}\!-\!\boldsymbol{x}^{*}}\!\!
		+\!\!\sqrt{\beta}\inprod{\boldsymbol{n}^k,\boldsymbol{x}^{k}\!-\!\boldsymbol{x}^{*}} \! \\ &\ \leq \sum_{i\in \mathcal{N} \setminus \mathcal{A}^k}\!\!\! \Big\lvert\inprod{\lambda_i\boldsymbol{g}_i\left(\boldsymbol{x}^k\right),\boldsymbol{x}^{k}\!-\!\boldsymbol{x}^{*}}\Big\rvert
		\!\!+\!\!\sqrt{\beta}\inprod{\boldsymbol{n}^k,\boldsymbol{x}^{k}-\boldsymbol{x}^{*}}\\
		 & \overset{(a)}{\leq} \sum_{i\in \mathcal{N} \setminus \mathcal{A}^k}\norm{\lambda_i \boldsymbol{g}_i\left(\boldsymbol{x}^k\right)}\norm{\boldsymbol{x}^k-\boldsymbol{x}^{*}}
		  + \sqrt{\beta}\inprod{\boldsymbol{n}^k,\boldsymbol{x}^{k}-\boldsymbol{x}^{*}}\\ & \overset{(b)}{\leq} RG\sum_{i\in \mathcal{N}}\mathbbm{1}_{\mathcal{N}\setminus\mathcal{A}^k}\left(i\right)+\sqrt{\beta}\inprod{\boldsymbol{n}^k,\boldsymbol{x}^{k}-\boldsymbol{x}^{*}},
	\end{aligned}
\end{equation}
where (a) follows from the Cauchy-Schwarz inequality, (b) from Assumption 1 and 2, while 
\begin{equation}  
	\mathbbm{1}_{\mathcal{S}}\left(i\right)= \begin{cases}
		1,   \quad \text{if} \quad i \in \mathcal{S}, \\
		0,  \quad  \text{otherwise}
	\end{cases}.
\end{equation}
denotes the indicator function. Finally, $I_3$ can be written as
\begin{equation}  
	\begin{aligned}\label{I3}
		 &I_3= \norm{\boldsymbol{\mathcal{L}}_x\left(\boldsymbol{x}^k,\boldsymbol{\lambda}^k\right)}^2+ \norm{\hat{\boldsymbol{\mathcal{L}}}_x\left(\boldsymbol{x}^k,\boldsymbol{\lambda}^k\right)}^2\\ &+2\inprod{\boldsymbol{\mathcal{L}}_x\left(\boldsymbol{x}^k,\boldsymbol{\lambda}^k\right),\hat{\boldsymbol{\mathcal{L}}}_x\left(\boldsymbol{x}^k,\boldsymbol{\lambda}^k\right)} =\norm{\boldsymbol{\mathcal{L}}_x\left(\boldsymbol{x}^k,\boldsymbol{\lambda}^k\right)}^2 \\ &+\norm{\sum_{i\in \mathcal{N} \setminus \mathcal{A}^k} \lambda_i \boldsymbol{g}_i\left(\boldsymbol{x}^k\right)}^2-2\sqrt{\beta} \inprod{ \boldsymbol{n}^k,\sum_{i\in \mathcal{N} \setminus \mathcal{A}^k} \lambda_i \boldsymbol{g}_i\left(\boldsymbol{x}^k\right)}\\
		&\!\!+\!\!\norm{\sqrt{\beta}\boldsymbol{n}^k}^2\!\!-\!\!2\inprod{\!\boldsymbol{\mathcal{L}}_x\left(\boldsymbol{x}^k,\!\!\boldsymbol{\lambda}^k\right),\sum_{i\in \mathcal{N} \setminus \mathcal{A}^k} \lambda_i \boldsymbol{g}_i\left(\boldsymbol{x}^k\right)\!\!-\!\!\sqrt{\beta}\boldsymbol{n}^k\!\!}.
	\end{aligned}
\end{equation}
Next, we take the expectation in both sides of \eqref{eq:p}. For term $I_2$ in \eqref{I2}, we have
\begin{equation}  
	\begin{aligned}\label{N-A}
		\mathbb{E}_{\mathcal{A}^k}\left[\sum_{i\in \mathcal{N}}\mathbbm{1}_{\mathcal{N}\setminus\mathcal{A}^k}\left(i\right)\right]&=\sum_{A=0}^N \mathrm{Pr}\left\{\lvert\mathcal{A}^k\rvert=A\right\} \left(N-A\right) \\
		&= N-\bar{A}_k\left(\beta\right),
	\end{aligned}
\end{equation}
where the expectation is taken w.r.t. the randomness of user participation, while $\bar{A}_k\left(\beta\right)\triangleq \sum_{A=0}^N A\cdot\mathrm{Pr}\left\{\lvert\mathcal{A}^k\rvert=A\right\},$ denotes the average number of users participating during the $k$-th round and
\begin{equation}  
	\mathrm{Pr}\left\{\lvert\mathcal{A}^k\rvert=A\right\}=\sum_{u=1}^{\binom{N}{A}}\bigg(\prod_{i \in \mathcal{A}_u}\gamma_i^k\left(\beta\right)\prod_{i \in \mathcal{N}\setminus \mathcal{A}_u}\left(1-\gamma_i^k\left(\beta\right)\right)\bigg),
\end{equation}  
where $\mathcal{A}_u$ is the $u$-th subset, among all $\binom{N}{A}$ subsets, with cardinality $A$. Furthermore, it is straightforward to show that the second term on the RHS of \eqref{I2} has zero expectation due to the zero mean of the AWGN, which leads to
\begin{equation}   \label{eq:R2}
	\begin{split}
		\mathbb{E}\left[I_2\right]&\leq RG\left(N-\bar{A}_k\left(\beta\right)\right). 
	\end{split}
\end{equation}
 Similarly, for term $I_3$ in \eqref{I3}, we have
\begin{equation}   \label{eq:EI3}
		\mathbb{E}\left[I_3\right]\leq L^2\left(\left(N-\bar{A}_k\left(\beta\right)\right)^2+1\right)+2LG\left(N-\bar{A}_k\left(\beta\right)\right)+\beta\sigma^2,
\end{equation}
where we have used Jensens' inequality by considering the convexity of $\norm{\cdot}^2$. By combining \eqref{I1}, \eqref{I2}, \eqref{I3}, \eqref{N-A}, and \eqref{eq:EI3}  we obtain Lemma 1a, which completes the proof.
\end{IEEEproof}
\emph{Lemma 1b:} We have
\begin{equation}  
	\begin{aligned}
		\mathbb{E}\left[\norm{\boldsymbol{\lambda}^{k+1}-\boldsymbol{\lambda}^{*}}^2\right]& \leq 
		\mathbb{E}\left[\norm{\boldsymbol{\lambda}^k-\boldsymbol{\lambda}^{*}}^2\right]+a_k^2L^2 \!\!\\ &+\!\!2a_k\mathbb{E}\left[\!\left(\!\mathcal{L}\left(\!\boldsymbol{x}^k,\boldsymbol{\lambda}^k\!\right)-\mathcal{L}\left(\!\boldsymbol{x}^k,\boldsymbol{\lambda}^{*}\!\right)\!\right)\!\right].
	\end{aligned}
\end{equation}
\begin{IEEEproof}
  The proof is similar to that of Lemma 1a, and thus, ommited due to space limitations.
\end{IEEEproof}
\emph{Lemma 2:} The following holds
\begin{equation}  
	\begin{aligned}
		&\mathbb{E}\left[\frac{\sum_{j=0}^{k-1}a_j\mathcal{L}\left(\boldsymbol{x}^j,\boldsymbol{\lambda}^j\right)}{Z_k}\right]-\mathbb{E}\left[\mathcal{L}\left(\boldsymbol{x}^{*},\boldsymbol{\lambda}^{*}\right)\right]\leq \\ & \frac{1}{Z_k}\Bigg(RG\sum_{j=0}^{k-1} \left(N-\bar{A}_j\left(\beta\right)\right)a_j +(\beta\sigma^2+L^2)\sum_{j=0}^{k-1}a_j^2 \\
		&  +2LG\sum_{j=0}^{k-1}\left(N-\bar{A}_j\left(\beta\right)\right)a_j^2+\mathbb{E}\left[\norm{\boldsymbol{x}_0-\boldsymbol{x}^{*}}^2\right]\\& +\sum_{j=0}^{k-1}L^2\left(N-\bar{A}_j\left(\beta\right)\right)^2a_j^2\Bigg).
	\end{aligned}
\end{equation}
\begin{IEEEproof}
	By rearranging the terms in \eqref{eq:lemma1a} and  consecutively adding both sides of the inequality for $j=0,1,...,k-1$, yields
	\begin{equation}  \label{adding}
		\begin{aligned}
			&2\mathbb{E}\left[\sum_{j=0}^{k-1}a_j\left(\mathcal{L}\left(\boldsymbol{x}^j,\boldsymbol{\lambda}^j\right)-\mathcal{L}\left(\boldsymbol{x}^{*},\boldsymbol{\lambda}^j\right)\right)\right]\\ &\leq  2RG\sum_{j=0}^{k-1}\left(N-\bar{A}_j\left(\beta\right)\right)a_j  +\sum_{j=0}^{k-1}L^2\left(N-\bar{A}_j\left(\beta\right)\right)^2a_j^2 \!\! \\
			&+2LG\sum_{j=0}^{k-1}\left(N-\bar{A}_j\left(\beta\right)\right) a_j^2+(\beta\sigma^2+L^2)\sum_{j=0}^{k-1} a_j^2\\ &+\mathbb{E}\left[\norm{\boldsymbol{x}_0-\boldsymbol{x}^{*}}^2\right]-\mathbb{E}\left[\norm{\boldsymbol{x}^{k}-\boldsymbol{x}^{*}}^2 \right].
		\end{aligned}
	\end{equation}
Since, the function $\mathcal{L}\left(\boldsymbol{x},\boldsymbol{\lambda}\right)$ is concave in $\boldsymbol{\lambda}$, for any fixed $\boldsymbol{x}\in\mathcal{X}$, it holds that
\begin{equation}   \label{lx}
	\mathcal{L}\left(\boldsymbol{x}^*,\hat{\boldsymbol{\lambda}}^k\right)\geq\frac{\sum_{j=0}^{k-1}a_j\mathcal{L}\left(\boldsymbol{x}^*,\boldsymbol{\lambda}^j\right)}{\sum_{j=0}^{k-1}a_j},
\end{equation}
Next, we divide both sides of \eqref{adding} by $Z_k=\sum_{j=0}^{k-1}a_j$ and combine it with \eqref{lx}. Finally, by utilizing the saddle-point theorem \cite{nedic2}, which states that for a saddle point  $\left(\boldsymbol{x}^{*},\boldsymbol{\lambda}^{*}\right)$, any $\boldsymbol{x}\in\mathcal{X}$, and for any $\boldsymbol{\lambda}\in \mathcal{D}$ the following holds
\begin{equation}  
	\mathcal{L}\left(\boldsymbol{x}^{*},\boldsymbol{\lambda}\right)\leq \mathcal{L}\left(\boldsymbol{x}^{*},\boldsymbol{\lambda}^{*}\right) \leq \mathcal{L}\left(\boldsymbol{x},\boldsymbol{\lambda}^{*}\right),
\end{equation}
the proof is completed. 
\end{IEEEproof}

Now, we are ready to prove Theorem 1. The rest of the proof can be conducted in a similar manner as the proofs of Corollary 1 and Lemma 2, as well as Proposition 5.1a in \cite{nedic2}.  One needs to  define $s=\sum_{j=0}^{k-1}a_j\boldsymbol{F}\left(\boldsymbol{x}^j\right)$, and take into account the convexity of 
$\boldsymbol{F}\left(\boldsymbol{x}\right)$, and Lemma 2. Afterwards, by following the same steps as the proof of Proposition 5.1a in \cite{nedic2}, it is straightforward to show that
\begin{equation}  
\begin{aligned}
&\mathbb{E}\left[ \norm{\left[\sum_{j=0}^{k-1}a_j \boldsymbol{F}\left(\boldsymbol{x}^j\right)\right]^{+}}\right] \leq \frac{1}{2r}\mathbb{E}\left[\underset{\boldsymbol{\lambda}\in\mathcal{D}}{\mathrm{max}}\norm{\boldsymbol{\lambda}_0-\boldsymbol{\lambda}^{*}}^2\right] \\ &+\frac{L^2}{2r}\sum_{j=0}^{k-1}a_j^2+\frac{1}{r}\mathbb{E}\left[\left(\sum_{j=0}^{k-1}a_j\mathcal{L}\left(\boldsymbol{x}^j,\boldsymbol{\lambda}^j\right)-f^{*}\right)\right]
\end{aligned}
\end{equation}
Then, by dividing both sides by $Z_k$, exploiting the convexity of functions $f_i\left(\boldsymbol{x}\right), \, i \in \mathcal{N}$, using Lemma 2, and the fact that $\boldsymbol{\lambda} \in \mathcal{D}$, the proof of Theorem 1 is completed.
\vspace{-0.2in}
\section*{Appendix B \\ Proof of Corollary 1}
Given $k\geq1$, we aim to minimize the RHS of \eqref{conviol} by properly selecting the value of $r$. Specifically, the optimal value $r^*$, satisfies
\begin{equation}  
r^{*}=\underset{r>0}{\arg\mathrm{min}}\left\{\frac{1}{r}\left(\delta_k\!+\!\frac{2}{Z_k}\left(\zeta\!+\!r\right)^2\right)\right\},
\end{equation} 
where $\delta_k$ and $\zeta$ are chosen such that the term $\frac{1}{r}\left(\delta_k\!+\!\frac{2}{Z_k}\left(\zeta\!+\!r\right)^2\right)$ is equal to the RHS of \eqref{conviol}. Hence, $\zeta\!=\!\frac{f_0\left(\bar{x}\right)-\tilde{q}}{\gamma}$ and $\delta_k$ is given in \eqref{eq:delta}. Since $r$ depends on the given round $k$, it is denoted as $r_k$, hereinafter. Next, in order to find the optimal value of $r^*_k$, we have 
\begin{equation}  
\begin{aligned}
&\frac{\partial}{\partial r^*_k}\left[\frac{1}{r^*_k}\left(\delta_k+\frac{2}{Z_k}\left(\zeta+r^*_k\right)^2\right)\right] = 0 \\ & \Rightarrow r^2 -\zeta r+2\zeta^2+\delta Z_k=0 \Rightarrow  r^{*}=\frac{\zeta+\sqrt{2\zeta^2+\delta_k Z_k}}{2}.
\end{aligned}
\end{equation}
Following that, the optimal projection set $\mathcal{D}_k$ in the $k$-th round is given by 
\begin{equation}  \label{setoptimal}
\mathcal{D}_k = \bigg\{\lambda \geq 0 \bigg| \norm{\lambda}_{\infty} \leq \frac{f_0\left(\bar{x}\right)-\tilde{q}}{\gamma} + r^{*}_k \bigg\}.
\end{equation}
By substituting $r^{*}_k$ in  \eqref{conviol}, we obtain
\begin{equation}  
\begin{aligned}
&\mathbb{E}\left[ \norm{\left[\boldsymbol{F}\left(\hat{x}_k\right)\right]^{+}}\right] \leq \\& \frac{2}{\zeta+\sqrt{2\zeta^2+\delta_k Z_k}}\left(\delta_k+\frac{3\zeta+\sqrt{2\zeta^2+\delta_k Z_k}}{2 Z_k} \right).
\end{aligned}
\end{equation}
Notice that $\delta_k$ contains terms of the form $\frac{\sum_{j=0}^{k-1}\left(N-\bar{A}_j\left(\beta\right)\right)a_j}{Z_k}$ and $\frac{\sum_{j=0}^{k-1}\left(N-\bar{A}_j\left(\beta\right)\right)a_j^2}{Z_k}$. These terms converge when $k \to \infty$. Specifically, $  
\frac{\sum_{j=0}^{k-1}\left(N-\bar{A}_j\left(\beta\right)\right)a_j}{Z_k} \leq N\frac{\sum_{j=0}^{k-1}a_j}{Z_k}=N
$ and
$\frac{\sum_{j=0}^{k-1}\left(N-\bar{A}_j\left(\beta\right)\right)a_j^2}{Z_k}\leq N\frac{L^2\sum_{j=0}^{k-1}a_j^2}{Z_k}.$
Therefore, since $  
\lim_{k \to \infty} N\frac{L^2\sum_{j=0}^{k-1}a_j^2}{Z_k} =0 \,\,\,$ we obtain $\,\, \lim_{k \to \infty} \frac{\sum_{j=0}^{k-1}\left(N-\bar{A}_k\left(\beta\right)\right)a_j^2}{Z_k} =0.$ It is concluded then, that $\lim_{k \to \infty} \delta_k < \infty$. Then, since $\lim_{k \to \infty} Z_k = \infty$, it is straightforward to show that 
\begin{equation}  
\lim_{k \to \infty} \mathbb{E}\left[ \norm{\left[\boldsymbol{F}\left(\hat{\boldsymbol{x}}^k\right)\right]^{+}}\right] =0.
\end{equation}
\section*{Appendix C \\ Proof of Theorem 2}
By considering the convexity of the objective function $f_0\left(\boldsymbol{x}\right)$ and taking into account that $\mathcal{L}( \boldsymbol{x}^*,\boldsymbol{\lambda}^*)=f^*_0$, we have
\begin{equation}  
\begin{aligned}
&\mathbb{E}\left[f_0\left(\hat{\boldsymbol{x}}^k\right)\right] \leq \mathbb{E}\left[\frac{\sum_{j=0}^{k-1}a_jf_0\left(\boldsymbol{x}^j\right)}{\sum_{j=0}^{k-1}a_j}\right]  \\ &= \mathbb{E}\left[\frac{\sum_{j=0}^{k-1}a_j\mathcal{L}\left(\boldsymbol{x}^j,\boldsymbol{\lambda}^j\right)-\sum_{j=0}^{k-1}a_j\inprod{\boldsymbol{\lambda}^j,\boldsymbol{F}\left(\boldsymbol{x}^j\right)}}{\sum_{j=0}^{k-1}a_j}\right].
\end{aligned}
\end{equation}
Thus, by using Lemma 2, we conclude that
\begin{equation}   \label{upbound}
\begin{aligned}
&\mathbb{E}\left[f_0\left(\hat{\boldsymbol{x}}^k\right)-f^{*}\right]\leq \!\! -\mathbb{E}\left[\frac{\sum_{j=0}^{k-1}a_j\inprod{\boldsymbol{\lambda}^j,\boldsymbol{F}\left(\boldsymbol{x}^j\right)}}{\sum_{j=0}^{k-1}a_j}\right]\!\!\\ &+\!\!\frac{R G}{Z_k}\Bigg(\!\!\sum_{j=0}^{k-1}\left(N\!\!-\!\!\bar{A}_j\left(\beta\right)\right)a_j\!+\!\mathbb{E}\left[\!\norm{\boldsymbol{x}_0\!-\!\boldsymbol{x}^{*}}^2\right]\!+\!\beta\sigma^2\sum_{j=0}^{k-1}a_j^2\\
& \!\!+\!\!2LG\sum_{j=0}^{k-1}\left(N\!\!-\!\!\bar{A}_j\left(\beta\right)\right)a_j^2\!+\!L^2\sum_{j=0}^{k-1}\left(\left(N\!-\!\bar{A}_j\left(\beta\right)\right)^2\!+\!1\right)a_j^2\Bigg).
\end{aligned}
\end{equation}
Next, we need to provide an upper bound for the term $-\mathbb{E}\left[\frac{\sum_{j=0}^{k-1}a_j\inprod{\boldsymbol{\lambda}^j,\boldsymbol{F}\left(\boldsymbol{x}^j\right)}}{\sum_{j=0}^{k-1}a_j}\right]$. Similarly to \cite{nedic2}, exploiting Lemma 1b and by taking into account that Lemma 1b holds for all $\boldsymbol{\lambda} \in \mathcal{D}$, we get 
\begin{equation}  
\mathbb{E}\left[\norm{\boldsymbol{\lambda}^{j+1}}^2\right] \leq \mathbb{E}\left[\norm{\boldsymbol{\lambda}^j}^2\right] +2\mathbb{E}\left[a_j\inprod{\boldsymbol{\lambda}^j,\boldsymbol{F}\left(\boldsymbol{x}^j\right)}\right]+a_j^2L^2. 
\end{equation}  
By adding both sides of inequality (79), for $j\!\!=\!\!0,...,k-1$, and then by dividing both sides with $\sum_{j=0}^{k-1}a_j$, we obtain the following
\begin{equation}   \label{eq:kati}
-\mathbb{E}\left[\frac{\sum_{j=0}^{k-1}a_j\inprod{\boldsymbol{\lambda}^j,\boldsymbol{F}\left(\boldsymbol{x}^j\right)}}{\sum_{j=0}^{k-1}a_j}\right] \leq \frac{\mathbb{E}\left[\norm{\boldsymbol{\lambda}_{0}}^2\right]}{2\sum_{j=0}^{k-1}a_j}+\frac{L^2\sum_{j=0}^{k-1}a_j^2}{2\sum_{j=0}^{k-1}a_j}
\end{equation}
Substituting  \eqref{eq:kati} in \eqref{upbound} and taking into account that $Z_k=\sum_{j=0}^{k-1}a_j$, the proof is completed.

\bibliographystyle{IEEEtran}
\bibliography{bib}
\end{document}